\documentclass[sigconf]{acmart}
\usepackage{multirow}
\usepackage[justification=centering]{caption}
\usepackage{framed}

\usepackage{booktabs} % For formal tables

\usepackage{graphicx}
\usepackage{subfig}

\begin{document}

\title{ Annotation Scaffolds for Object Modeling and Manipulation }

\author{Pablo Frank-Bolton}

\affiliation{%
  \institution{The George Washington University}
  \streetaddress{800 22nd St NW, Suite 4000}
  \city{Washington} 
  \state{DC} 
  \postcode{20052}
}
\email{pfrank@gwu.edu}

\author{Rahul Simha}

\affiliation{%
  \institution{The George Washington University}
  \streetaddress{800 22nd St NW, Suite 4000}
  \city{Washington} 
  \state{DC} 
  \postcode{20052}
}
\email{simha@gwu.edu}

\begin{abstract}

We present and evaluate an approach for human-in-the-loop specification of shape reconstruction with annotations for basic robot-object interactions. Our method is based on the idea of model annotation: the addition of simple cues to an underlying object model to specify shape and delineate a simple task. 
The goal is to explore reducing the complexity of CAD-like interfaces so that novice users can quickly recover an object's shape and describe a manipulation task that is then carried out by a robot.
The object modeling and interaction annotation capabilities are tested with a user study and compared against results obtained using existing approaches. 
The approach has been analyzed using a variety of shape comparison, grasping, and manipulation metrics, and tested with the PR2 robot platform, where it was shown to be successful.

\end{abstract}


\maketitle

\section{Introduction}

Embedding robots in social contexts often requires human participation in terms of decision and control. 
Novel situations, difficult decisions, and human preference are sure to impose new conditions to a robot's working environment, further complicating its perception, decision, and action process.
While fully autonomous robots 
remain out of reach, there is a clear need for appropriate ways for humans to control or influence robot actions.
Human-in-the-loop (HIL) 
is the term used in the modeling of semi-automatic systems in which humans are used as an integral part of the design  to gather data, for decision-making, or guidance.

If a human is to interact with a robot, an effective means of communication must be established through a properly designed interface. While many different types of interfaces abound, a two-dimensional Graphical User Interface (GUI) is often a necessary method of interacting with the robot and the environment in which it functions. Visualization in 2D and an interface involving Mouse-Monitor-Keyboard (MMK) is still the most used and least expensive of interfaces for Human-Computer Interaction (HCI).

While alternatives to the traditional GUIs have been proven effective, they often require special equipment, intensive training, or the use of non-intuitive methods of control. The simple 2-dimensional feedback and the tried and tested traditional input methods as well as the newer touch and gesture-based commands have been successfully applied in robotics \cite{Rouanet2013,singh2013interface}.

Once a need for human participation has been identified, there still remains the question of the level of control over the robot's decision process.
Complete supervision through remote-control allows users to fully specify robot actions, reducing the need for autonomous decision-making and therefore simplifying the system's implementation. This can be helpful when safe teleoperation is desired, like with bomb disposal \cite{yamauchi2004packbot} or exploration of dangerous wrecks \cite{akal2004application}. While extremely useful, this type of control can be highly demanding of the human operator, and might require intensive training.

As an alternative to full control, human operators may provide guidance through the use of \emph{annotation}. Annotation refers to the inclusion of supplementary information to a dataset (or a robot's perceptual stream) 
that, when properly interpreted, produces a relatively large amount of additional semantic information. 
Examples include the addition of cues for object segmentation \cite{wang2012touchcut,Sengupta2013}
and the indication waypoints for motion \cite{Osentoski2010}. These hints, in the form of annotations, are most advantageous when they are effective while requiring as little effort as possible on the part of the annotator. Ideally, one expects high \textit{usability} and a low \textit{cognitive load}. 
Usability refers to the annotation mechanism's ease of learning and use, while cognitive-load refers to the mental effort in accomplishing some annotation task.  
The use of annotations for robot learning and operation is a rich and growing field \cite{Wang2011,kostavelis2015semantic}. If properly designed, interfaces that allow the gathering of such annotations can help streamline robot control, as well as a valuable source of data for future autonomous systems.

One problem with designing a system with a human in the loop is that the human can make errors. Another is that a particular designated human may not be available when most needed.
In terms of availability and effort, crowdsourcing offers the infrastructure to distribute the load among a host of annotators. This scheme has been successfully applied to other  tasks like image segmentation and object recognition, and to 
help robots grasp novel objects or navigate through unknown environments \cite{Osentoski2010,Kehoe2015,DeCroon2014}.
In addition, one can leverage the wisdom of crowds to extract an additional level of information about the multitude of annotation instances \cite{ prelec2017solution, hu2014learning}. In essence, one can look at each human cue as an opinion, which could function as a simple voting system for tallying preference. 
A knowledge-base of annotations would then contain solutions for instances and, when considered as a whole, an implicit overview of different human approaches to each problem.

As mentioned above, one benefit of these annotations is that their use might exceed the purpose for which they were originally created. As long as the relevant information is stored in a way that could be exploited for solving a different challenge, reuse is a possibility.  
The reference frame for annotation can be the actor (like a robot), or the object (like a door). When annotating with the object as the reference point, 
the instructions address the modes of use, rather than the user's motions. One immediate benefit of this is that object-relative annotations may be readily reused, as long as the appropriate measures are taken to allow different actors to participate.
One way of creating objects for reuse is through their design and modeling using Computer-Assisted-Design (CAD).
Under this approach, real or invented object models are created to represent shape, or to indicate the inner workings of a mechanism. These approach can be then extended by adding information that could indicate other modes of use. CAD, however, is a difficult discipline to learn, which indicates the need for simpler object-relative annotation strategies. 

First, for an annotation system to be effective, it must capture as much of
the user's intentions as possible with the least amount of effort 
on their part. To accomplish this, the interface feedback and 
controls need to be designed to
maximize human expression while minimizing cognitive load. Since user intentions vary, versatility in task annotation is desirable, that is: it should provide many ways to solve the same task, and also ways to address different tasks. 

Second, the annotation system should be designed with casual users
in mind, as opposed to requiring intensive training. A system
that requires significant training is unlikely to find application
in consumer robotics.

Third, the annotations must have impact on the target application. 
In our application of 
robotics, the annotations must be useful to accomplishing a 
robotics task that is considered hard for autonomous robots.

Fourth, an annotation system should feature the ability to store
annotations for repeated use in the future (this is what makes it superior
to continuous teleoperation), and to aggregate across the
annotations of multiple users. Learning to construct these stored annotations should be easier than learning traditional CAD.

Our research is focused on the design and evaluation of a simple annotation strategy
that allows HIL 
for two classic tasks in robotics:
object modeling, and Pick-and-Place. 
These tasks are fundamental building blocks of a wide range of high-level tasks that a service robot is expected to encounter.

\section{Related Work}

\subsection {Teleoperation and Simple GUI Control}

Remote control may be accomplished through the use of a wide variety of %controlling 
interfaces \cite{boboc2012review}.
These have changed considerably with the advent of Human-Centered Design (HCD)
and have evolved from joystick-and-button black boxes to force-feedback devices \cite{farkhatdinov2010preliminary}, smart gloves \cite{Ekvall2007}, body-suits \cite{Ramos2015}, verbal instructions \cite{Johnson-Roberson2011}, and vision-based analysis of natural body motions \cite{lipton2018baxter,pollard2002adapting,fritsche2015first,ishiguro2017bipedal}. While attractive, an immediate complication in using these methods is obtaining and learning to use the required equipment. In addition, mixed results from immersive teleoperation have been reported, resulting in efforts to reduce the cognitive load of the operators \cite{hart1988development}. In \cite{martins2015design} a clear trade-off was detected: greater control and situational awareness came at the cost of possible cognitive overload and impaired performance. 

On the other hand, there has been a lot of effort in developing methods for reducing the cognitive load on the user in 2D-GUIs. In \cite{kent2017comparison}, simple constraining of motions was effective in offloading some of the cognitive effort. The familiarity and widespread use of smartphones and tablets has also brought attention to touch-based control. In \cite{singh2013interface}, a simple touch-based interface was aimed at reducing operator fatigue.

Interfaces based on Monitor, Mouse, and Keyboard are a simple and cheap alternative that has been applied to a wide variety of robotics applications like robot navigation \cite{Osentoski2010}, grasping \cite{miller2004graspit,Sorokin2010,sucan2013moveit}, or object manipulation \cite{leeper2012strategies,Sung2015}. 

\subsection {Interface Design}

Work on human perception points to the fact that we process 3D objects as arrangements of 2D views  \cite{koenderink1979internal,bulthoff1995three}. In addition, several tasks involving shape understanding only need 2D views, like recognition and detection \cite{cyr2004similarity}. 

Display type may play a part on the ability to perform different tasks. In the work done by St John et al \cite{st2001use} a distinction is made between scene understanding and specific tasks involving the precise judgment of relative position. They compared pure 2D interface without projective effects to 3D displays: 2D views of 3D scenes with perspective effects. They reason that, while 3D displays seem compelling, integrated and natural, they can cause ambiguity or distortion by the nature of their presentation in a 2D display. In particular, for tasks requiring the precise relative position of two objects, 2D views proved superior. 

On the other hand, Tory et al \cite{tory2006visualization} argue that small cues, like shadows, can be added to pure 3D displays to improve approximate navigation and relative position. In addition, shape understanding is greatly improved by the presence of perspective views of objects. They went on to test hybrid displays (3D with 2D capabilities) in tasks involving shape understanding and relative position, and concluded that combination displays did better than strict 2D or 3D displays. They determined that perspective views help with 3D integrated spatial tasks while 2D views help with precise actions requiring concentration and the understanding of relative position.

\subsection {Annotation}

Informative \emph{annotation} can be obtained from users in several ways. One possibility is to add tags and qualifiers to a dataset after it has been acquired. This can range from intensive human labeling in post-production to methods that require minimal markings.
Full image segments are drawn by humans and used as ground truth in \cite{martin2001database}, whereas in \cite{wang2012touchcut}, the segmentation algorithm needs only an initial seed point provided by a human.  
In 3D environments, human-segmented data is often used for 
training and validation \cite{rodola2013scale}. Simple hints have also been used in these environments to seed automatic segmentation algorithms \cite{nguyen2017robust}. Annotations can also be used for tracking objects in 3D from an initial seed labeling \cite{Teichman2013}. In \cite{zivkovic2008sensors}, an annotated dataset is constructed for robot navigation.

Learning from Demonstration (LfD) offers an alternative to the active addition of annotations in post-processing. 
In LfD, the parameters of actuation are implicitly refined through the observation of examples provided by humans. This happens because under this strategy,
perception-action systems are built specifically to follow human actions and replicate them within their own circumstances (like different kinematic structures or safety constraints). See \cite{argall2009survey} for a thorough review.

A third approach is \emph{kinesthetic teaching},
where an operator indicates appropriate motions by physically guiding the robot while it records the event \cite{akgun2012trajectories}. Alternatively, the robot may be guided virtually, aided with a variety of feedback options \cite{ruffaldi2017vibrotactile, restrepo2017iterative}.

In one sense, annotation can be obtained without human supervision by automatically processing the data and extracting useful intelligence from the collected information. Metadata can be analyzed as a form of annotation for the payload it describes. Often, the input data itself comes from data sources rich with explicit or implicit human annotation. In \cite{patron2012structured}, automatic classification methods are trained using existing human-interaction datasets that have been previously grouped by activity. 
While effective in its own right, this type of unsupervised learning needs to be validated before being put to used in any context where safety, preference, or informed judgment is required. One problem with unintentional annotation is that it is unstructured and noisy by nature, making the extraction of usable intelligence a considerably more difficult proposition \cite{fu2012attribute}. When possible, it is preferable to use annotations that are directly aimed at enhancing the understanding of a situation or task that one intends to refine.

\subsubsection {Annotation of Perceptual Data}
There are different mediums for annotation that the robot may provide or use as input for operation. When using kinesthetic learning, the system focuses on the recording of changes in its joint space with respect to a given objective \cite{akgun2012trajectories}. While useful, it is usually more powerful to also have a way of capturing the surrounding environment. A common approach is to use visual perception of some kind. This is due to the enormous amount of relevant information that may be extracted from this approach. It may capture color, changes in spatial and temporal relations, and importantly, it's a perceptual stream  that we as humans generally favor to analyze our environments. Annotation of images has a long tradition in robotics. For traditional image stills or streams of images (2D), large databases have been constructed for segmentation \cite{martin2001database} and object recognition \cite{caltechImageDatabase,mitImageDatabase}. More recently, stereo vision and depth sensors have extended the perceptual stream to 3D. In this realm, several annotated databases have been constructed \cite{3dWarehouse,chang2015shapenet}. The volumetric and color information has been used for object recognition \cite{Xiang2016}, tracking \cite{Teichman2013}, or shape reconstruction \cite{nguyen2017robust}.

Off-the-shelf sensors like the Microsoft Kinect give easy access to these input streams, and have made the 3D scanning of environments and objects ubiquitous. Several algorithms have been developed and refined to allow the integration of depth sensing onto unified models of the scenes they capture \cite{Han2013}. The Kinect Fusion algorithm \cite{newcombe2011kinectfusion}, and its many variants \cite{whelan2012kintinuous}, allow the tracking of a scene and integration of captured depth-frames into a coherent volumetric representation. Some variants have even accomplished deformation and part-tracking \cite{dou20153d}. One important aspect of the models that these approaches create is their sensitivity to the sensor and the context. The Kinect has a resolution of about 2 mm and has trouble capturing thin surfaces, transparent or shiny objects and requires careful scanning or large amounts of computation for dealing with noise. While various techniques exist to deal with these problems, when these appear in conjunction to context-dependent constraints, human annotations become highly valuable. For an in-depth review of automatic and semi-automatic methods of reconstruction from point clouds, see \cite{berger2017survey}.

\subsubsection {Annotations and Crowdsourcing}

Annotations that help solve particular instances of problems can be gathered into a knowledge-base that can, in turn, be studied to gain insights into the solving of more general situations. 
One way to gather a large amount of usable annotations is to take advantage of the growing infrastructure for \emph{crowdsourcing} tasks, like Amazon's Mechanical Turk \cite{buhrmester2011amazon}. 

The use of the public for the annotation of robotic tasks imposes a series of additional conditions on the annotation interface as well as the level of required user effort, expertise, and precision. Simple methods must be developed for non-experts to participate. Existing technology can simply use human guidance to refine world models. In \cite{Chen2015Maps}, crowdsourcing is used to construct maps by tracking the motion of smartphone recordings of indoor environments.
One immediate requirement is that the necessary input device be readily available and preferably of intuitive use and design. That is why point-and-click or touch-screen interfaces have been so widely used for this purpose. In \cite{nguyen2017robust}, a robust annotation tool is provided to segment captured 3D data. In \cite{Sorokin2010}, crowdsourcing was used to segment, classify and evaluate 2D and 3D objects for grasping. A series of computer vision problems are addressed using human computation in \cite{gingold2012micro}. See \cite{Kehoe2015Survey} for a comprehensive survey of cloud robotics, which includes several applications using crowdsourcing.
In \cite{hu2014learning}, multiple labels are used to combine annotations in a simple voting scheme that can help deal with inconsistent or erroneous labeling. Advances in automatic methods and annotation have developed into powerful hybrid methods that can achieve a variety of tasks combining the best of both worlds \cite{russakovsky2015best}.

\subsubsection {Annotation and Cognitive Load}

One important aspect of GUI design is to consider the user and the objectives. Several studies have looked at what humans can and can't do in these robotics tasks, what interaction exists between an individual's abilities, and what the interface can do \cite{hart1988development,Oviatt2006}. In addition, studies involving tasks in 3D space usually consider human natural abilities or previous experience in similar environments \cite{leeper2012strategies}. Indeed, studies suggest that spatial reasoning ability not only varies between individuals, but that different tasks might require different abilities within the realm of spatial reasoning \cite{vandenberg1978mental,Blazhenkova2010}. One common precept in HCD is to avoid visual clutter in order to lower cognitive load. Studies have found that novice users get better results on interfaces that hide extraneous features \cite{Reis2012}, and where they only focus on the high-level tasks while the robot controls the low level details \cite{Enes2010}.  

\subsection {Task-Oriented Annotation}

In the field of annotations for robotic tasks, there has been intense research under both the \textit{sense-plan-act}, and \textit{behavior-oriented} approaches. On one hand, there is perception-stream annotations for 2D and 3D data. 

\subsubsection {Annotation for Object Reconstruction}

Object reconstruction refers to the recovery of a real object's structure in digital form from some stream of perceptual data. 
Significant progress has been made in 
automatic reconstruction. One of the general methods consists of using multiple 2D views of an object to obtain its shape \cite{dyer2001volumetric,moons20093d}. While this approach allows the use of monocular cameras, depth-annotated images have become easily accessible and allow for accurate pose estimation and refinement of models \cite{Sun2013}. The work of Newcombe et al \cite{newcombe2011kinectfusion} used the Microsoft Kinect sensor to integrate depth-augmented images into a single volumetric model of the scene. This approach was very successful and has been used and refined extensively \cite{whelan2012kintinuous,dou20153d,Zhang2014}. 

The main problem with automatic reconstruction is its sensitivity to input resolution and any noise or artifacts the sensor  might cause. A well known problem has to do with the natural difficulties in dealing with thin structures, noisy scans, or shiny or transparent surfaces \cite{ummenhofer2013point}. Several algorithms attempt to combat these problems by imposing correction mechanisms to preserve shape or inferred structure \cite{Sharf2007,li2011globfit,lafarge2013surface}.

An alternative to automation is full manual reconstruction or object modeling. Examples of this approach can be found under the category of CAD.
Several programs of varying complexity have been developed to construct 3D shapes and have been used in digital object construction and animation \cite{solidworks,3DsMax}. These programs are usually quite powerful but prohibitively complex for crowdsourcing. Learning them is also quite difficult, with basic courses ranging from a few weeks to a couple months. 

A third option is semi-automatic reconstruction. Here, the idea is to use human cues to seed or refine automatic methods. These can work on 2D images, like the work of Chen et al\cite{CZSHC-2013}, where simple human marks on 2D images may result in the creation of 3D objects. The work of Kholgade et al\cite{kholgade20143d} integrates simple annotations on 2D images with 3D object models extracted from a repository to extract complex objects from 2D scenes.
Another approach is to work directly with 3D scans of a scene. This can either be on the depth-based point clouds or the reconstructed mesh. In Arikan et al\cite{arikan2013}, simple cues help snap planar primitives on top of point cloud sections in order to build simple architectural models. In the work of Nguyen et al\cite{nguyen2017robust}, annotations are made on reconstructed 3D models of the whole environment to generate segmented 3D scenes. In the work of Sharf et al \cite{Sharf2007}, topology-aware reconstruction is first attempted automatically, and then iteratively corrected with human input. A refined interface showing a similar approach is shown in the work by Yin et al \cite{Yin2014}.

One important distinction between methods is the form these annotations take. In \cite{nguyen2017robust,CZSHC-2013} simple lines and sketches are interpreted as instructions and combined with the section of the images they are used on. The results are either 3D models of the 2D-segmented shapes, or a segmented 3D space, where each part of the scene may be labeled with the object categories. Another typical objective is the creation of new object models from existing ones. One approach is to deform them into new shapes. In the work of \cite{Nan2010,zheng2012interactive}, cuboid proxies are used to guide shape deformations. 

The idea of using shape proxies for future actions over the base shape is, in a sense, a method of annotation. In the work of McCrae et al\cite{mccrae2011slices}, a slice-based proxy was developed for representing 3D meshes. These were placed by humans and later used as training input for an automatic planar-proxy creation algorithm. The potential applications range from printing simplified 3D objects to the annotation and recovery of 3D shapes. In a follow-up work, McCrae et al \cite{mccrae2013surface} used these planar proxies, in conjunction with crowdsourcing to study user abilities regarding estimation of surface normals in commonly occurring 3D object models.

\begin{figure}[!htbp]
\centering
\includegraphics[width=3in]{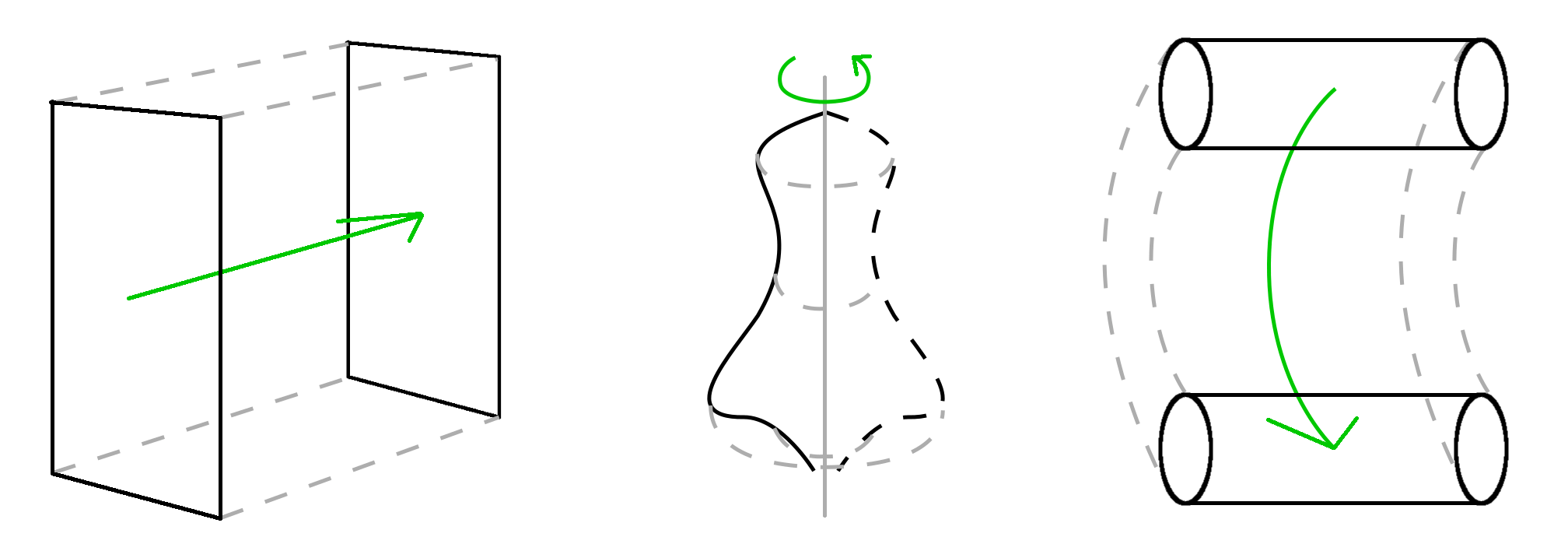}
\caption{Examples of swept surfaces that generate closed volumes}
\label{fig:sweptSurfs}
\end{figure}

A \emph{swept surface} is the area resulting from moving a line, which can be closed (contour) along a path. These can be translational, rotational or a combination of both. 
They can even apply to solids.
A \emph{generalized cylinder} \cite{binford1975visual} is a type of translational swept surface that parametrizes the characteristics of the surface depending on its position along the sweep. 
If the resulting surface encloses a volume, then the resulting shape can represent a solid in 3D. Some examples of shapes that can be created this way are shown in Figure \ref{fig:sweptSurfs}. For the purposes of this work, all mentions of sweep-based objects refer to  generalized cylinders.

In the work of Yoon et al \cite{yoon2006sweep}, a \emph{sweep-based proxy} is extracted from 3D object meshes and later used to deform them by performing simple modifications of the underlying proxy. This approach highlights the power of simplified representations for object analysis and modification. The work of Shtof et al \cite{Shtof2013} shows the use of these simple sweep-based proxies to fit 2D sketches and transform them to 3D objects.

Yin et al \cite{Yin2014}, developed an annotation scheme that uses simple gestures and the structure of object point clouds to infer a set of underlying part proxies that can be modified to reconstruct the shape. This gesture-based reconstruction approach can help deal with incomplete scans and was designed specifically to offer the best trade-off between user-effort and shape quality. The inference of underlying shape is carried out automatically, with some parameter tweaking done by the operator.  One disadvantage is that this automatic reconstruction has trouble dealing with multiple small holes or with topologies that contain cavities, like mugs or bowls. A similar approach has been used to segment meshes into generalized cylinders \cite{Zhou2015}. Two typical topology-related problems are objects missing cavities, like a mug without a cavity \cite{Yin2014} or unexpected topologies like mugs with various holes \cite{mccrae2011slices}.

Sweep-based object representations have a long history in the study of shape and perception. They have also been called generalized cylinders \cite{naeverepresenting}, and together with \textit{implicit volumes}, 
\textit{constructive-solid-geometry} (CSG) 
and \textit{boundary representation}, constitutes one of the main strategies for representing shapes. Part of their appeal is that they are intuitive, simple, easy to generalize and scale into more complex shapes. It also allows many parameters of the resulting solid to be easily calculated \cite{ballard1982cm}. They have also relatively straightforward conversions to surface representation and shape skeletons \cite{huang2013l1}, which are useful alternatives when dealing with 3D objects.

\subsubsection {Annotation for Object Grasping}

Robotic grasping has been attempted automatically in several ways. These approaches vary in the level of dependency on existing information repositories.

Some studies avoid using object models and use the perceptual input directly to extract grasp affordances \cite{piater2011learning}. In \cite{popovic2011grasping}, local surface features are used to propose grasps. Some techniques use a multitude of heuristics to generate grasp hypotheses from the sensed point clouds of the scene \cite{hsiao2010contact}. These approaches are, in a way, model-less and allow the grasping of objects by their current appearance. An alternative approach uses object models to store and seed the inference of grasps. In the work of \cite{Ciocarlie2007,ciocarlie2009hand}, Ciocarlie et al., develop a technique called \textit{Eigengrasps} where low-dimensional grasp subspaces are computed that match the object shape. These in turn may be used for seeding interactive grasping approaches. Some approaches are denominated hybrid, since they use appearance features or 3D-model detection depending on the situation \cite{Brook2011}. While automatic grasping is a promising field, context and functional requirements might place additional conditions on the choosing fo a specific grasp.

A slightly less ``independent'' procedure is to match and apply grasping templates to the sensed input \cite{herzog2012template}. For graping based on existing knowledge-bases, one option is to detect known objects followed by executing a specific grasp linked to that particular instance or that category of objects \cite{dang2012semantic,azad2007stereo}. In \cite{Miller2003}, shape primitives are linked to the target object and used to seed appropriate contact-level grasping. 

Databases of grasping examples may be used as a starting point for automatically generating a grasp hypotheses \cite{li2005shape}. While these approaches are not fully automatic, they already contain important semantic information that might be germane to a multitude of context-dependent grasping tasks. In \cite{dang2012semantic,nikandrova2015category}, shape and objective constraints are used to specify appropriate grasps. 

Humans are excellent at quickly arriving at functional grasps, even for novel objects. We have the advantage of context and experience to help guide our choice. Whether for direct grasp suggestions, or for creating labeled model data, human annotations have been widely used. Human action may be observed and used under the LfD strategy and then processed to generate robot grasps \cite{romero2009modeling}. Alternatively, human input can serve to label grasping databases or to interactively choose grasps \cite{ciocarlie2009hand}. In summary, a range of automatic and interactive methods exist that can use either raw perception and automation or existing models to obtain grasps. For a survey on data-driven grasp synthesis, see \cite{Bohg2013}. 

Teleoperated grasps can be done combining two general methods: using a GUI to indicate waypoints \cite{sucan2013moveit}, and contact-reactive mechanisms to refine the mechanical aspects of that particular grasp \cite{hsiao2010contact}. Force-feedback can help the user get a sense of the grasp resistance and object malleability during the interaction \cite{khurshid2017effects}.

Significant effort has been placed on obtaining a taxonomy of types of grasps and to judge their physical qualities \cite{Cutkosky1989,feix2016grasp,liu2016annotating}. These studies draw inspiration from early analysis of human grasping \cite{napier1956prehensile} or robot hands \cite{li1989grasping}. One widely used approach for judging grasp quality is that of the metric developed by \cite{ferrari1992planning}. It uses the grasp contact points to construct  a convex hull over the set of wrenches that can be applied with the given grasp. This is called the % GWS.
Grasp Wrench Space (GWS). 
Two measurements that can be used are the volume of the computed convex hull and the radius of the maximum sphere contained in it, which we call the epsilon-distance, or $\epsilon$-dist.
These indicate the robustness of a grasp and the 
versatility of the 
wrenches applicable for the given grasp. One popular tool used in the literature to evaluate grasp quality and generate grasp candidates is GraspIt! \cite{miller2004graspit}. It uses the GWS volume and $\epsilon$ metrics to rank grasps that have been generated using a variety of grasp-planning techniques including the above mentioned \textit{Eignegrasps}.
\\

While several methods abound that use existing 3D mesh objects or that detect them from 2D-views or point clouds, less effort has been paid to linking grasps to simpler shape representations or proxies like the ones obtained in \cite{yoon2006sweep,mccrae2011slices}. As mentioned previously, these proxies are compact versions of objects, and can be used to represent categories as well as instances, and help indicate shape, as well as topology. Linking grasps to these might work as a sort of look-up table for grasps.

\subsubsection {Annotation for Object Handling}

Grasping and manipulation are closely related. For some grasps finding the final contacts of the gripper is insufficient. In several tasks, the approach to accomplish the grasp, as well as the task-dependent manipulation that will be performed affect the generated grasp hypotheses. Some studies have looked at shape category and the physical context to determine manipulation constraints \cite{nikandrova2015category}, while others have used human experience to seed approach vectors \cite{Ekvall2007}. In \cite{dang2012semantic}, a grasp planning algorithm is used in conjunction to a set of task-dependent semantic constraints to choose grasps, which in turn may inform the approach vector. These semantic constraints are extracted from example grasps and then used to construct a semantic affordance map which directly relates the object class (obtained from object depth information) to the approach vectors and different task-appropriate grasps.  This provides evidence that relating object instance or category models to manipulation annotations is of great utility.

In terms of obtaining grasp and manipulation examples, annotated databases or human-examples may be used. Several studies have focused on using human input to learn or validate manipulation tasks. The work by Kent et al \cite{Kent2014} uses \emph{crowdsourcing} of manipulation tasks to reconstruct 3D object models (as integrated point clouds) with grasp points attached to them. This approach demonstrates the power of crowdsourcing techniques to complex tasks in robotics. It also represents an example of semantically-augmented object models that can be later used for solving tasks.  

One popular tool for motion planning is Moveit! \cite{sucan2013moveit}. This tool can be used within the larger Robot Operating System (ROS) to perform automatic motion planning with collision avoidance. This may also be used in conjunction with a manipulation interface that humans can use to indicate motions. In the work by Leeper et al, \cite{leeper2012strategies}, this interface was used to analyze manipulation strategies for indicating grasps and grasp approach methodologies. They compared continuous teleoperation against three other strategies with varying degrees of autonomy. They showed than a combination of manual annotation and automatic processing resulted in the greatest success rate. This is supported by the work done by Hertkorn \cite{hertkorn2016shared}, where the importance of shared workload is thoroughly analyzed in the context of grasping and manipulation.

\subsubsection {Annotation Usability Studies}

We based our object modeling usability study on
similar studies involving interactive shape reconstruction from point clouds \cite{Sharf2007,Nan2010,arikan2013}. These constitute the state-of-the-art on this type of interactive approaches \cite{Berger2014,berger2017survey}.
In these studies, user interactions with their software are logged and timed, and later analyzed for precision errors. In \cite{Sharf2007,Nan2010}, an expert user illustrates the possibilities of their interface, while in \cite{arikan2013}, five users are tested. Similarly to these studies, we recorded action choices and timing and evaluated shape quality. We also chose to include an expert user to show the possibilities of the interface if enough time is devoted to learning it.

In the area of object handling, the work by Hertkorn \cite{hertkorn2016shared} has one study with a similar objective (evaluating the effect a particular grasping assistance technique). In this study, $20$ participants were were asked to complete $30$ grasp trials each. Basic shapes were chosen to attempt to eliminate the effects of shape complexity on the interface's effectiveness in assisted grasping. Similarly, we chose to use simple and/or familiar basic shapes.

Weisz et al \cite{weisz2013user} uses a database of preplanned grasps, as well as an online planner to help a user find appropriate grasps. Time and success rate was computed for five subjects attempting three tasks. While five subjects is a small number, results indicate trends and allow the research to hone in on the appropriate refinements for a more in-depth evaluation. 

The work by Leeper et al \cite{leeper2012strategies} is designed to compare different interaction strategies for attaining valid grasps using the PR2 robot. In this study, $48$ participants were asked to grasp as many objects as possible (between $2$ and $9$) in three rounds of trials. Participants were shown a tutorial before the start of the trial.

In the work of Sorokin et al \cite{Sorokin2010}, several tasks and trials were tested using crowdsourcing. Here, the annotation tasks were simple in order to allow testing the approach, and later integrated with the ultimate objective of completing pick-and-place tasks. They found that in many cases a single average ``worker'' produced poor results, but that by cleaning and averaging $3-5$ of them, high accuracy was obtained. 

The work by Rouanet et al \cite{Rouanet2013} uses a simple protocol that involves input and output surveys, a tutorial, and a challenge section. While they were focused on evaluating different methods of directing robot motion and attention, we used the same protocol structure for our own annotation tests. 

\section{Methods and Experimental Design \label{sec:methods}}

In this section, we describe the design rationale behind the %\gls{PCS}
Point Cloud Scaffolds (PCS)
and the 
Point Cloud Prototyper (PCP).
In addition, we describe the user study used to evaluate the annotation interface and subject's abilities in the tasks of object modeling, grasping, and pick-and-place.

\subsection {Point Cloud Scaffolds}

\subsubsection {Getting Point Clouds}

Given the availability of popular and inexpensive depth-sensing equipment, we decided to base our reconstruction effort on RGB-D technology.
The Point Cloud Library (PCL) 
\cite{Rusu_ICRA2011_PCL} provides a large set of methods and a well maintained API for working with point clouds. It seamlessly integrates with the Microsoft Kinect, perhaps the most popular sensor in its class, and has a large community behind it. 

The Kinect sensor can take snapshots of a scene at a resolution of 640 by 480 pixels. These %\gls{RGB} 
Red-Green-Blue (RGB) images can be registered to a depth map in a way that causes each pixel have three color dimensions (r,g,b) plus three spatial dimensions (x,y,z). While extremely rich, this raw input needs to be processed and corrected to overcome the natural noise and resolution limitations of the sensor. In particular, registration is not perfect, which means that the color image might not be overlaid exactly on top of the depth map, causing some pixels to contain the wrong color or and for others to have no depth. This is a consequence of the mechanism for obtaining the depth, which uses an RGB camera and an 
Infra-Red (IR) emitter-receiver pair at very close - but different - locations. The fact that the Kinect uses an IR emitter and sensor to capture depth, also means that it is sensitive to scene characteristics affecting this approach. Specular reflections, like those present in shiny objects impede depth sensing. Also, objects made of glass or clear plastic are not detected properly. Resolution is also an issue, since the Kinect can capture differences down to only about $2 mm$, which means that thin structures like nets, cables, or leafs are often invisible.  

A single snapshot can capture only what it can see from one point of view (POV), which means that only parts of the scene are captured. This can be called $2\frac{1}{2}$-D. While a good  
Red-Green-Blue-Depth (RGB-D) snapshot might be enough for a human to understand and interpret, the missing information makes the automatic recovery of the object structure a difficult task. One option is to capture multiple snapshots from different points of view and then registering (coalescing) them offline. This is where the Kinect Fusion approach proves extremely useful. PCL contains a version of the Kinect Fusion algorithm from Newcombe et al \cite{newcombe2011kinectfusion}, called \textit{Kinfu}. Kinfu uses each depth-frame to correct a running occupancy model of the scene by assigning an occupancy weight to the voxelized environment. The only requirements are that there is sufficient spatial texture in the scene and that the scanning is done in a relatively smooth way. Spatial texture means non-smooth changes in the geometry of the scene. This is needed for the algorithm to have clearly identifiable characteristics that can be matched to the current view. Smooth scanning is needed because jumps in the position or the POV make the current view and the running model too different to match quickly. The matching algorithm, called
Iterative Closest Point (ICP), 
needs the structures to be matched to start in relatively close positions to avoid falling into local minima. In the end, a voxelized space is iteratively refined to carve out areas that are not occupied. This model can be converted to mesh or point cloud form and used to represent the scanned space. While Kinfu is extremely powerful, it cannot model occluded areas of the scene, like internal structures or hard-to-reach spaces in the environment. In addition to the above mentioned issues with noise and resolution, object models are often incomplete or noisy. This is where human annotation comes in.

The idea is simple: Scan a scene and extract a point cloud of the objects of interest; then ``trace'' the desired shape over this initial scan to refine and model the virtual object. The concept of \textit{tracing} refers to the one used to replicate 2D images, where one can place a translucent piece of paper over the original design and draw over the important features. While the resulting image might contain less details than the original, the \textit{tracer} does not need any of the expertise from the original creator and can focus on exactly the details they might consider relevant. The challenge is that this approach needs to be migrated to 3D, where the point cloud can serve as the ``original'' and the object is to come up with an equivalent to the translucent paper to recreate the shape. One approach is to simply use projections of the shape into 2D-planes that can then be analyzed to extract the desired contours. The problem with this approach is that cavities and internal structures will not be visible. This, however, is a good starting point for the design of a 3D trancing scheme. In the following section, we relate the steps taken to construct a tracing overlay, which we call a %\glsdesc{PCS}, 
Point Cloud Scaffold, to accumulate important contours and capture possible cavities and internal structures.

\subsubsection {Scaffolds for 3D Tracing}

A tracing overlay for point clouds needs to be a simple structure that can collect as much (or as little) information as the tracer chooses. It must also follow some intuitive tracing mechanism that can work for a variety of shapes and levels of exactitude. As mentioned above, contour sweeps have a long history in computer science and are understood to be versatile, yet simple and intuitive structures. Complex or organic looking shapes may be created  
using a contour swept along a path fitted over a 3D shape, like in the work of Zhou et al \cite{Zhou2015}.

\begin{figure}[htp]
\subfloat[][]{\includegraphics[width=3in]{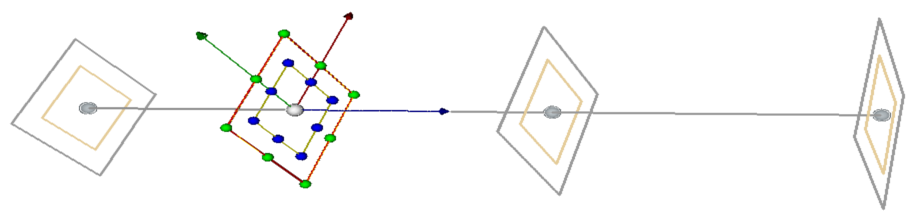}\label{fig:SCA_long} }\\
\subfloat[][]{\includegraphics[width=0.6in]{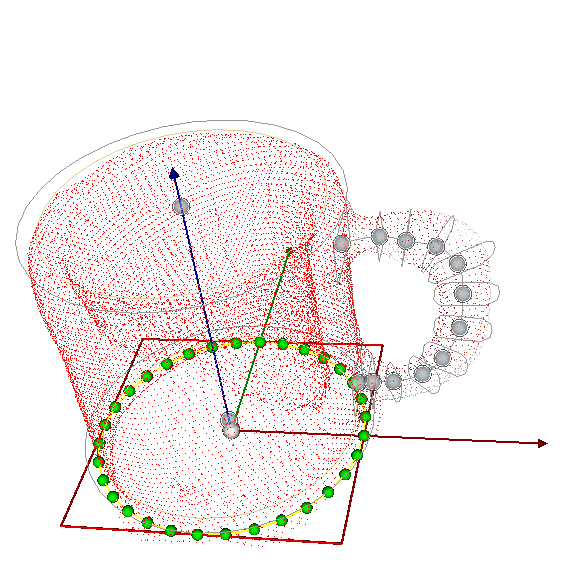}\label{fig:SCA_mug_a} }\hfill
\subfloat[][]{\includegraphics[width=0.6in]{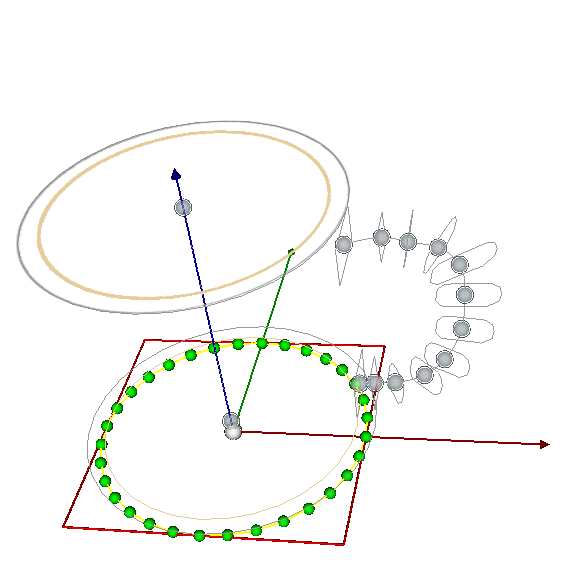} \label{fig:SCA_mug_b} }\hfill
\subfloat[][]{\includegraphics[width=0.6in]{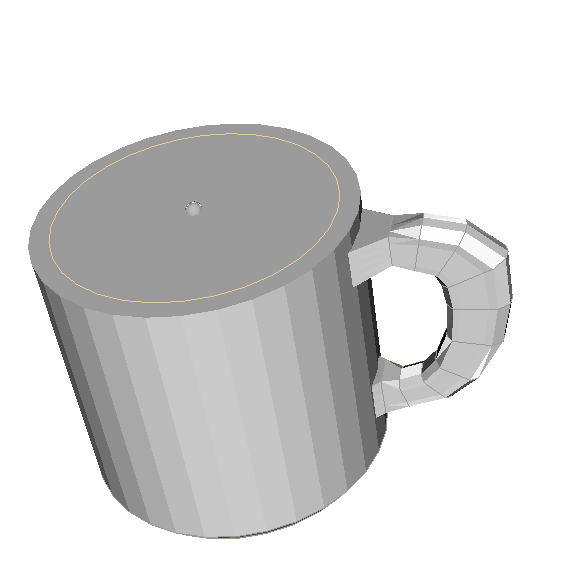}\label{fig:SCA_mug_c} }\hfill
\subfloat[][]{\includegraphics[width=0.6in]{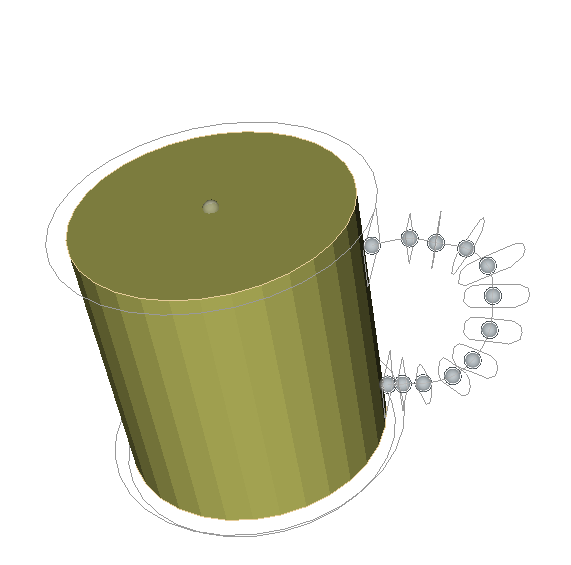}\label{fig:SCA_mug_d} }\hfill
\subfloat[][]{\includegraphics[width=0.6in]{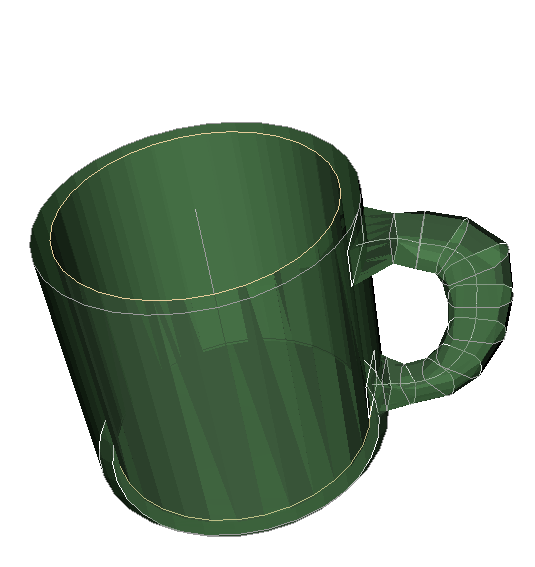}\label{fig:SCA_mug_e} }
\caption{ Scaffolds:  \protect\subref{fig:SCA_long} trivial scaffold. In the selected slice plane, green handles define contour vertices and blue handles define hole vertices (external contours shown in silver and hole contours shown in gold); \protect\subref{fig:SCA_mug_a} scaffolds over point cloud; \protect\subref{fig:SCA_mug_b} scaffolds for cup and handle; \protect\subref{fig:SCA_mug_c} external mesh; \protect\subref{fig:SCA_mug_d} hole mesh; \protect\subref{fig:SCA_mug_e} combined mesh.}
\label{fig:SCA_mug}
\end{figure}

A swept surface is the result of integrating into a volume the space occupied by a planar shape as it moves through it. A circle swept along a straight line creates a cylinder, while a square would create a prism. This approach can be extended by allowing the swept shapes to change scale, shape, or even to contain internal holes. There are several research avenues that use this approach to describe or reconstruct shapes \cite{schroeder1994implicit,osher2001level,Abdel-Malek2006,Yin2014,Zhou2015}. The description of such a structure can be done explicitly (by placing structural features that control the shape and the path of the contours), or implicitly (by constructing some sort of analytical description that can be queried at any point of interest). Given its intended use (human editing) we chose to describe the sweeping structure explicitly through contour and path features. Figure 
\ref{fig:SCA_mug}  shows the parts of our sweep based structure, which we call: a %\gls{PCS}
Point Cloud Scaffold.

To describe a contour, we used a closed interpolating spline using a Cardinal basis \cite{schoenberg1973cardinal}. This types of objects describe lines in space that can be defined piece-wise and controlled by a set of points over their domain. This spline is described over a plane and then transformed to occupy any location in space. In our implementation, we control the spline shapes using a series of control points (also on the plane) that can either form part or be outside of the spline itself. Since we want the points not only to control the shape, but also to form part of it, we use a form of splines that interpolate these points into a closed and continuous curve. 
In Figure \ref{fig:SCA_long}, the (red square) contour, as well as the (green sphere) control points can be seen.

Once two or more of these contours are defined on different planes, these can be connected in a sequence to form a complete scaffold. To form the volume, the contours must connect to those before and after them until the whole object is closed. We used a simplified linear interpolation between consecutive contours, which allows very simple implementation and can be quickly refined into complex shapes. Figure %\ref{fig:scaff} 
\ref{fig:SCA_mug_c} shows the closed structure resulting from joining the contour planes of the two different scaffolds, which can be seen by themselves in Figure \ref{fig:SCA_mug_b}. 

A possible extension for extending the topology of the initial volumes is to allow the path to be closed, meaning that the swept contour returns to itself to form a closed path. This would easily allow the creation of tori or similar objects. 

Finally, we included the possibility of adding a second closed spline to every plane. This second spline constitutes a planar hole that can also be connected to contiguous holes to create volumetric negative spaces. This simple scheme allows the creation of basic internal structures that would otherwise remain unspecified if automatic POV-based reconstruction was used by itself. This mechanism can also be extended to include more internal complexity by adding alternating layers of positive and negative spaces. Figure 
\ref{fig:SCA_mug_d} shows the whole (negative) volume. The positive and negative volumes can be intersected (as in Constructive Solid Geometry) to generate a complex final structure with cavities (Figure \ref{fig:SCA_mug_e}).

All methods for interacting with the scaffolds will be described in the prototyping module of the description of PCP.
It is worth noting that the use of point clouds as a basis for the sweep-based scaffolds is for the implementation of PCP as a proof of concept. The idea of sweep-based annotation scaffolds is potentially extensible to other 3D spatial occupancy representations like the one in the work of Wurm et al \cite{Wurm2010}.

\subsection {PCP: The Point Cloud Prototyper}

\subsubsection {PCL Interface Design}

We use the tracing analogy as a method of constraining user interaction to a reduced set of operations. Some tracing actions require focused attention on the position of parts and the distance relations between objects. 
From previous research \cite{st2001use,tory2006visualization}, it is known that 2D views are better for these types of position-dependent tasks, while 3D perspective views are preferable when dealing with high level scene understanding. We therefore utilized a type of interface that can accommodate these two types of actions. For general scene visualization and navigation, 3D perspective views are used. For precision tasks, users were able to work either under orthographic views, or directly constrained to edit scaffold elements in a plane.

\subsubsection {PCL Implementation}
PCP is based on an early project from the Point Cloud Library called ``Cloud Composer'' \cite{cloudComposer} developed by J\'er\'emie Papon \cite{papon2012}.

PCP has three main modules: point cloud editing and visualization (which was the purpose of the original cloud composer from PCL); object modeling using scaffolds; and manipulation using a gripper widget and a joint mover. 
In this work, we describe the object modeling and manipulation modules.

\begin{figure}[!htbp]
\centering
\includegraphics[width=3.2in]{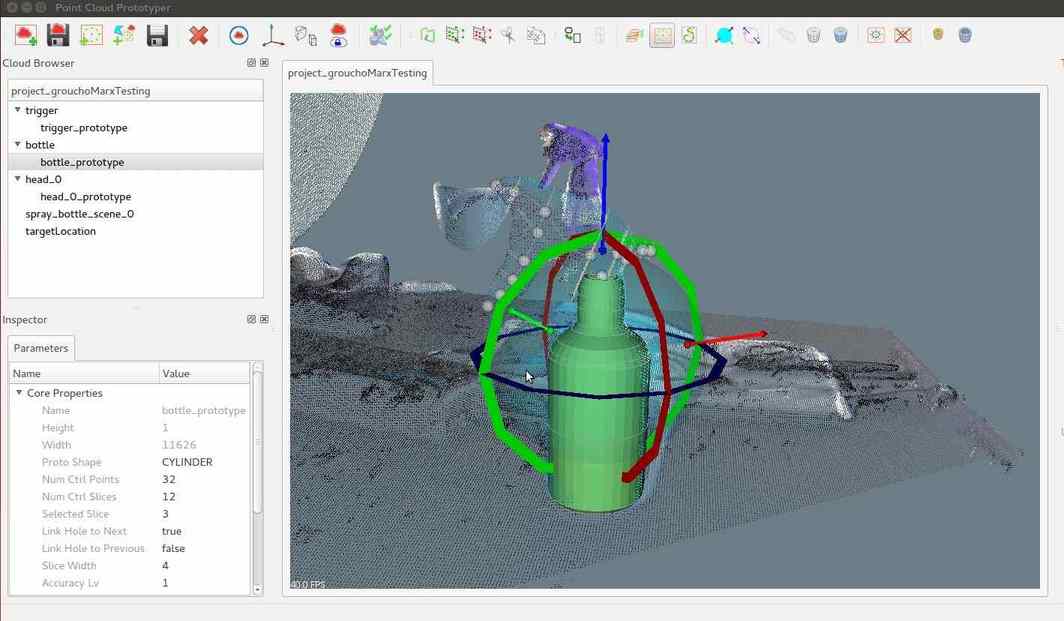}
\caption{The PCP GUI during the construction of a spray bottle. }
\label{fig:PCP}
\end{figure}

\subsubsection {Point Cloud Visualization and Editing}

The Cloud Composer project uses PCL's \textit{PCLVisualizer} to display and keep track of point clouds and meshes. It is also built in a way that allows the inclusion of plugins that may serve as filters or processing tools for point clouds. These include filters like cloud sanitizing and voxel grid downsampling; feature estimation like normal extraction; and segmentation actions like euclidean clustering and supervoxel extraction. While these filters are useful and can be integrated into the shape reconstruction workflow, we chose to refrain from their inclusion since we wanted to see what could be accomplished from user tracing over raw input clouds.

\begin{figure}
\centering
\subfloat[][]{\includegraphics[height=0.8in]{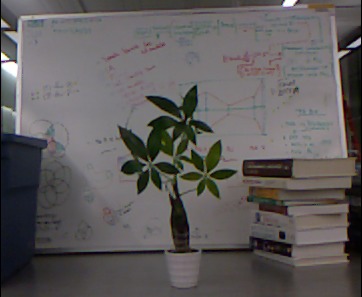}\label{fig:plantRGB}}
\hfill
\subfloat[][]{\includegraphics[height=0.8in]{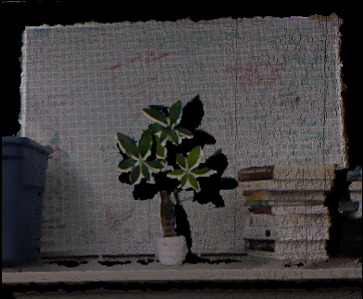}\label{fig:plantRGBD}}
\hfill
\subfloat[][]{\includegraphics[height=0.8in]{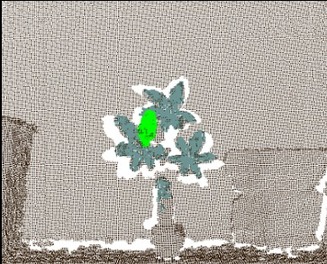}\label{fig:plantSegmented}}
\caption{RGB image of a scene \protect\subref{fig:plantRGB}; point cloud of the scene \protect\subref{fig:plantRGBD}; Segmented cloud \protect\subref{fig:plantSegmented}}
\label{fig:plant}
\end{figure}

Cloud Composer has a basic interface for applying the filters and editing the clouds. We made several modifications to the existing interface to improve clarity and interaction. One important addition to this module was the inclusion of a polygonal point selection tool to select parts of point clouds. This allowed users to precisely segment a scene into areas and objects into parts, each of which can be modeled independently. In addition, we included methods for hiding point clouds or preventing their modifications, which allowed users to have visual markers for understanding the scene, without running the risk of having visual clutter or causing unintentional changes. While RGB-D frames carry the scene's colors, we also added the option to paint clouds differently for added clarity. Lastly, we performed several minor changes to the way some actions were called or the details of their execution. For example, we added buttons for centering the view of clouds, or switching the projection from perspective to orthographic. Figure \ref{fig:plant} shows the sequence of steps to scan a scene Figure \ref{fig:plantRGB} to obtain a cloud Figure \ref{fig:plantRGBD} , and segment the resulting cloud using the point cloud editing tools Figure \ref{fig:plantSegmented}.

\subsection{Annotations for Reconstruction} \label{sec:Annotations for Reconstruction}

\subsubsection {Inserting Scaffolds}

The first step was adding the scaffolds as new editable items. We followed a hierarchical approach where the full \textit{scaffold} is composed of planar contours called \textit{slices} which are in turn determined by control points or \textit{handles}. As this tool is focused on reconstruction from point clouds, each scaffold needs to be directly connected to an underlying cloud.
This is easily modifiable in case parts need to be added which, for some reason, have no point clouds to be attached to. This, however, was not a requirement since we wished to reconstruct shapes that were only mildly incomplete or had holes. While this \textit{point-less} modeling extends the capabilities of the system, it moves more closely towards the approach used in CAD, which we did not want to emulate.
 
Two methods were tested for fitting PCS to point clouds:
insertion through oriented bounding box (OBB), and POV-insertion through crosshairs. OBB-insertion first needs to obtain an oriented bounding box, which is a bounding box where the three orthogonal directions run along the principal directions of change in the point cloud. This can be easily computed using PCL, and used to place a scaffold. The scaffold could be placed in such a way that the \textit{slices} ran along the axis with the greatest change. After insertion is performed, a user may change the direction of the scaffold (switch the direction the sweep follows), or permute the position of the scaffold inside the box (switching the direction of sweep among the three main axes of the bounding box). While this approach proved successful in fitting regular and elongated shapes like long cylinders or prisms (like a baseball bat), it proved impractical when dealing with more compact shapes (like an apple).

\begin{figure}[!htbp]
\centering
\includegraphics[width=3in]{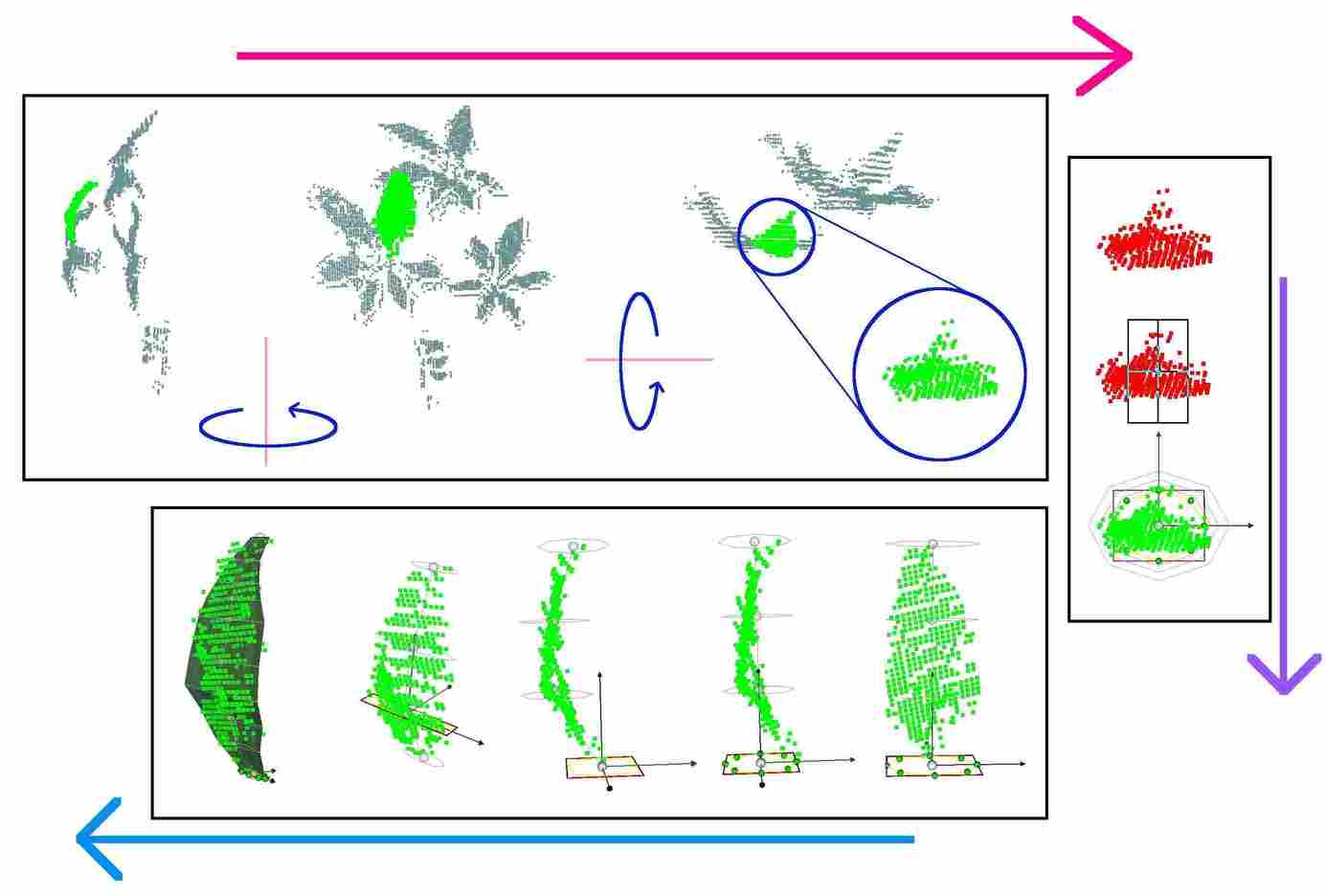}
\caption{Shape selection shown with respect to the whole object (pink segment), scaffold insertion (purple segment), and editing (blue segment). }
\label{fig:modelingAndInsertion}
\end{figure}

POV-insertion follows the idea that a user can choose and quickly identify the sweep direction for a shape. For instance, a user could quickly pick the sweep direction for both a baseball bat and an apple regardless of the length of the shapes. The steps are: $1$) the user places the view along the direction of the sweep, $2$) chooses one of two scaffold primitives (cylinder or box), and $3$) executes the scaffold insertion. This can be done in three quick actions with minimal delay. The insertion process analyzes the structure of the point cloud to refine the positioning of the scaffold around it. It will place the first and last contours at sufficient distance to exactly cover the nearest and furthest points, and will change the area of the cross section (the slice widths and heights) to encompass the chosen point cloud.  Figure \ref{fig:modelingAndInsertion} shows the whole process of segmenting the point cloud from Figure \ref{fig:plant}, setting the POV, inserting the scaffold, and editing its shape to generate a curved object.

\subsubsection {Editing Scaffolds}

As mentioned above, several tasks involving 3D tracing require focus on positions and distances, for which we provided embedded interaction modes in 2D. As previous research has found \cite{kent2017comparison}, constraining interactions can help focus users and lower cognitive load.

\paragraph{Editing the slices:} the contours in each slice can be independently translated, rotated and scaled using an arrow and disk transform widget that is highlighted for each slice. 

\paragraph{Editing the Sweep-Axis:} One action involves modifying the path along which the surfaces are swept. We call this the \textit{sweep axis}, or SA. While a straight shape (like a baseball bat) does not need a curved path, there are several shapes that do (like a banana). For this reason, the SA must be editable. The path that the slices follow can be manually placed in 3D by displacing every slice independently using the independent slice controls.

While useful, the independent placement of many slices can be slow. This is why we included an \textit{axis-drawing widget}. When activated, the widget places a restriction on the movement that the slice centers can follow, which reduces their placement to a problem in 2D.  While currently not needed for this work, this widget can be easily extended by including a free-hand drawing capability for the path.

\paragraph{Editing the number of parts:} The initial number of handles per slice and slices per scaffold can be modified. This is done so in a properties tab. One additional method for altering the number of slices is through their direct insertion (or deletion) over the sweep-axis.

\paragraph{Editing handles:} Handle positions can be individually repositioned by dragging them along their constraining plane. Several easy refinements are possible by imposing shape patterns over individual handle contours, like polygons, or rectangles. We also allow the copying and pasting of handle arrangements from other slices.

\paragraph{Editing holes:} Any slice can be extended with a planar hole. This is done so by simply clicking a button that inserts an internal closed spline, also controlled through handles. For the hole to extend to 3D, another planar hole must be added to a neighboring slice. We've included the possibility of applying certain actions to multiple/all slices at a time, which would allow the simultaneous addition of planar holes, quickly creating volumetric ones. Hole handles can be edited in the same way as the external contour ones and can be independently scaled.

\paragraph{Editing the whole scaffold:} The scaffold as a whole can be translated, rotated and scaled using a scaffold-wide arrow and disk control widget. A secondary scaling feature allows the distance between slices to be altered. In addition, holes can be independently scaled.

\paragraph{Shrink-Wrap mode:} The above editing possibilities can be helpful when precise control is necessary, but may take some time to complete. For that reason, a shrink-wrap feature has been added that allows contours to "wrap around" the underlying point cloud for quickly resetting the handles and the position of the slice centers. This, in combination with the manual editing allows fast modeling of organic/oblong shapes that would otherwise demand more effort and attention to complete manually.

\subsubsection {Visualizing the Prototype}
The running object reconstruction can be visualized in several ways. One is to show the external contour-based mesh, which we call the \textit{skin} (Figure \ref{fig:SCA_mug_c}). One can also show the internal \textit{hole mesh} independently. This will show the ``negative'' volume created by the hole sequences (Figure \ref{fig:SCA_mug_d}). A combination of the contour and internal meshes produces a \textit{difference mesh}, which is the result of removing the negative volume from the contour one. The \textit{final mesh} visualization displays a union of all the different difference meshes for each of the object parts (Figure \ref{fig:SCA_mug_e}). Two other possible visualizations are \textit{wireframe} and \textit{point cloud} views. All of these visualizations are constructed from joining contiguous internal and hole contours through their handles. These can all be exported and saved to file.

\subsection {Annotations for Object Grasping for Pick-and-Place}

An additional annotation that can be informed by the object's shape (scaffold) is the motion of a gripper that manipulates it. While the precise grip force and attitude
can be refined automatically, the high-level decision of where to grasp and how to manipulate an object
can be left to the operator. Once a location is chosen, grasping can be refined automatically based on appearance and tactile feedback \cite{hsiao2010contact}, the detection and fitting to predefined models \cite{dang2012semantic}, or a combination of both \cite{Brook2011}. Many such approaches
even use human instructors to help a robot infer axes and ranges of motion \cite{Sung2015}.

For annotating manipulation, we added a gripper module that the subject may use to record virtual gripper poses with respect to the object and saved as waypoints. These waypoints constitute a sequence of poses for a gripper to follow. These poses have six Degrees-of-Freedom (DoF) and are composed of a position in space: $x,y,z$ and an orientation, which can be specified by a quaternion: $w,x,y,z$ (quaternions have four coordinates but only 3 degrees of freedom).  In our case, we used a simplified PR2 parallel gripper since we used this robot to test manipulation.
 
Several features where added to simplify or speed-up the movement of the gripper in space. Two types of scene navigation are possible: egocentric (where the POV is from the agent being moved, the gripper in this case), or exocentric (the POV is from an external perspective). Previous work \cite{ferland2009egocentric} has found that both modes have different benefits, and that providing both capabilities, users gain a better perspective of the environment.
The basic exocentric manipulation is done using arrows and disks similar to the ones used in \cite{leeper2012strategies,miller2004graspit}. We also provided an egocentric POV for the gripper. This is very similar to the views used in First-Person-Shooter (FPS) video games for maneuvering in a scene. 

\subsubsection {Annotating Grasps}

The gripper itself is represented using a simple widget composed of the arrows and disks as well as the usable contact planes, which are the interior of fingers and palm. While other work \cite{leeper2012strategies,miller2004graspit} has used full meshes to represent the gripper, we wished to see if this basic representation is a valid alternative. This gripper may also be opened and closed to show the user how contact would occur.

We included a grasp evaluation mechanism to test the validity of a candidate grasp. This mechanism uses the \textit{GraspIt!} API developed by Jennifer Buehler \cite{buehler2015}. It computes the volume and 
$\epsilon$-distance 
of the
Grasp Wrench Space (GWS)
obtained from the chosen grasp. This can help the user verify the validity of a grasp or decide between different grasp alternatives.    

\begin{figure}[!htbp]
\centering
\includegraphics[width=2in]{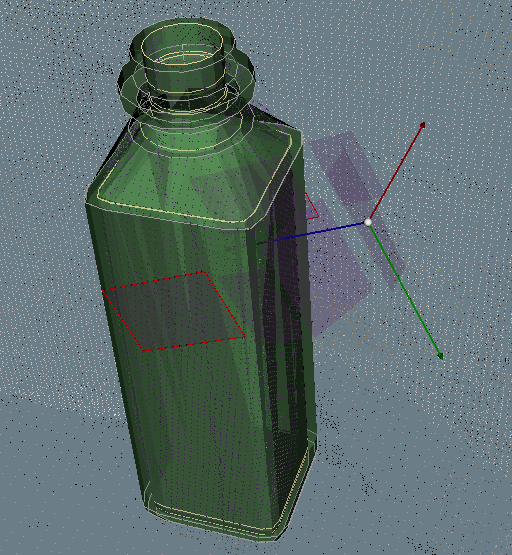}
\caption{Grasping a reconstructed bottle using PCP. The gripper is represented using a simplified widget composed of contact surfaces. }
\label{fig:PCPGrasp}
\end{figure}

The grasp annotation itself is the 6 DoF pose indicated by the user for the object grasp. An added feature is the inclusion of the pre-grasp, which represents a possible approach vector for the current scenario. This annotation can be saved with respect to the robot or with respect to the object scaffold, depending on the desired scenario. For the first case, we used the point cloud coordinate origin as the robot's POV, since we wished to localize the scene with respect to a head-mounted Kinect Sensor. For the second case, the origin of the scaffold was set to the center of the first slice defining the object scaffold, 
(the base slice). Figure \ref{fig:PCPGrasp} shows a grasp indication using PCP. 

\subsubsection {Annotating Rigid Object Handling}

The waypoints that indicate motion are stored as 6 DoF poses containing the following sequence: a pre-pose, a grasp-pose, and any number of handling waypoints. The scaffold can be used to easily create a representation of the object for visualizing its manipulation. We use a mechanism where a user can see a \textit{ghost} version of the object being displaced by the gripper. This allowed users to take the object dimensions into account while displacing it in space.

As mentioned above, these annotations are task dependent, and therefore require a higher level semantic mapping between objects and tasks. While this is beyond the scope of this thesis, a possible implementation could relate object scaffolds through waypoint annotations. As an example: two object scaffolds, a milk carton and a bowl, could be connected through a grasp and waypoint-sequence to indicate the action "pour". Although this path would depend on the spatial context, a multitude of these examples would constitute a rich dataset for extracting high-level understanding of the complex ``pour'' task.

\subsection{Experimental Design} \label{sec:Experimental Design}

There are two important aspects in evaluating the scaffold-based annotation scheme. The first involves measuring the effectiveness of the designed implementation for accomplishing tasks and the ease with which subjects can learn to operate it. The second has to do with analyzing the characteristics of the interaction itself in terms of what users can do well, and what constitutes a limitation for valuable annotation. To obtain metrics for both aspects, the following protocols were designed to contain ways of measuring interaction and annotation quality as well as objective (quantitative) characteristics of the annotation process. 

We split the usability study into two main modules: object modeling from point clouds, and pick-and-place. The object pick-and-place module was, itself, divided into two separate tasks: grasping of objects, and handling of gripped objects in space.
In order to evaluate the interface's precision and ease of use, we compare it to alternative software and interaction schemes that might serve as a ``gold standard''. 

We made a separate user study (with different sets of users) for each of the following experiments:

\begin{itemize}
\item PCP Object Modeling Experiment, where subjects used  our interface to construct object scaffolds from pointclouds.
\item PR2 Pick-and-Place Experiment, where subjects guided the PR2's robot arm and gripper through a set of Pick-and-Place tasks, similar to \cite{Balasubramanian2014}.
\item PCP Pick-and-Place Experiment, where subjects were asked to complete a set of Pick-and-Place tasks using PCP.
\end{itemize}

\paragraph {Preliminary Subject Assessment}

Every user was asked to complete entry and exit surveys as well as a spatial reasoning evaluation based on the Mental Rotation Test (MRT) \cite{vandenberg1978mental}. The entry survey was used to record self reported measures of previous experience with different categories of UIs. The exit survey was used to inquire about perceived effort and difficulty in completing their assigned tasks. The spatial reasoning evaluation is a reduced version of the MRT that contains 12 of the 24 questions. 
We used the analysis of the (MRT) by Caissie et al \cite{caissie2009does} to choose the sequence of questions that would maintain the general structure of the test while reducing its duration. 

The surveys were self reported values in the range $[1,5]$ and the spatial reasoning evaluation score, or (SR) was an integer in the range $[0,12]$, with $12$ being the highest possible score.

\paragraph {Participants}

Participants were contacted from within the university through mailing lists and flyers. For the reconstruction experiment we had $14$ participants, $6$ male, and $8$ female, with $13$ in the $18-22$ age range and one of $27$; in the PCP pick-and-place experiment, we had $16$ participants, $8$ male, and $8$ female, with all in the $18-24$ age range;  in the PR2 pick-and-place experiment, we had $18$ participants, $12$ male, and $6$ female, with $15$ in the $18-22$ age range and two with $35$ and $41$ years.

\subsubsection {User Study: PCP Object Modeling}

The high level evaluation approach was to compare the resulting reconstructed shapes obtained from PCP to alternative methods. The alternative methods included an automatic reconstruction method: Kinfu; and different solution based on CAD: Solidworks \cite{solidworks}. 

\paragraph{Experimental Setup (PCP):} 

Participants use a normal keyboard and mouse setup in front of a screen. They first must complete a tutorial that uses videos and asks them to follow along with PCP. Once they have completed this stage, thy are shown each of the tasks and asked to compete them using PCP. No time constraints were imposed on them. One important aspect of the approach is that users were allowed to see the objects they were supposed to model. This can be justified by the fact that CAD users had access to the original objects and Kinfu is based on a scan of the original objects as well. Users may instead see snapshots of the objects instead of having direct access to them, which would be sufficient to help them counteract the deficiencies present in the scanned clouds.

\paragraph{Protocol:}

For reconstruction, subjects were asked to follow a tutorial that would teach them to use PCL. The tutorial lasted between 1 and 2 hours and was designed to illustrate all of the reconstruction features that we wished to evaluate. While this approach does not directly lend itself to the crowdsourcing scenario, it can be greatly simplified and refined for that purpose. The idea of showing the full set of features was intended to allow users to display the action methodologies that best suited them. This would, in turn allow us to better evaluate user capabilities and the program itself. 

After the tutorial was completed, subjects were asked to complete two reconstruction stages. In the first, they were given the task of reconstructing four shapes. They were told that they could take up to 40 minutes to complete a shape, and were encouraged to try different approaches and tools. In the second stage, they had up to 20 minutes to reconstruct a different set of four objects. The objective of the two stages was to evaluate any change in object quality due to time restrictions. We also wanted to compare error metrics between stages for a pair of objects of the same category.

\subparagraph{Choice of Objects:}

The objects were chosen to include shape characteristics that would require the use of different techniques provided by PCP. In recent surveys on the state-of-the art of reconstruction from point clouds,  Berger et al \cite{Berger2014,berger2017survey} observed that the classes of shapes that were reconstructed fell into the following categories: 

\begin{itemize}
\item CAD models: these refer to assemblies of simple geometric primitives. 
\item Organic shapes: free-form, often curvilinear structure.
\item Synthetic shapes: shapes with canonical geometric properties and regularity.
\item Architectural models: A subset of Synthetic shapes, usually contain a great deal of global regularity and functional constraints.
\item Urban environments: Large scenarios composed of a limited number of object types. Also contain several functional constraints and regularities.
\item Indoor environments: Mixture of Synthetic and organic shapes, also with a reduced number of object types.
\end{itemize}
 
\begin{figure}
\centering
\subfloat[][]{\includegraphics[height=0.7in]{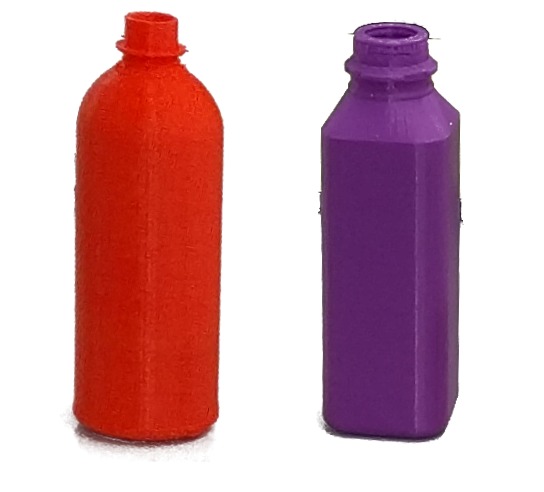}\label{fig:bottles}}
\hfill
\subfloat[][]{\includegraphics[height=0.7in]{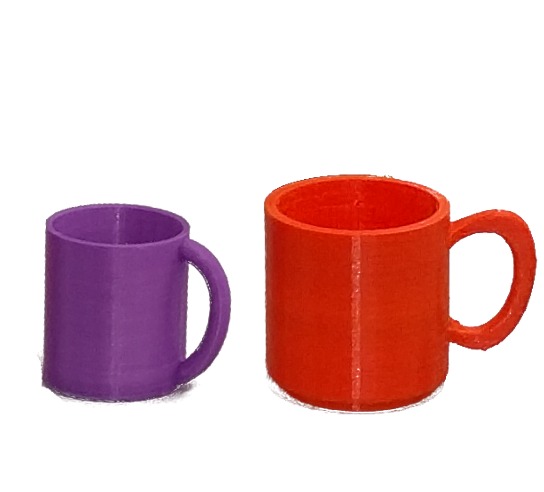}\label{fig:mugs}}
\hfill
\subfloat[][]{\includegraphics[height=0.7in]{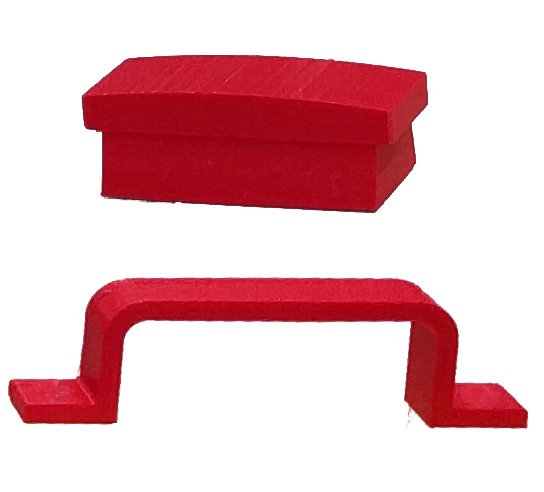}\label{fig:handles}}
\hfill
\subfloat[][]{\includegraphics[height=0.7in]{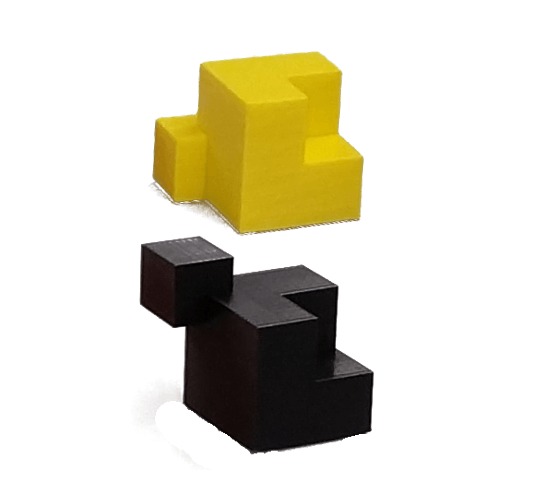}\label{fig:cuboids}}
\caption{
\protect\subref{fig:bottles} Bottles (Synthetic objects) ; 
\protect\subref{fig:mugs} Mugs (Synthetic with organic parts) ; 
\protect\subref{fig:handles} Handles (Synthetic with different functions) ; 
\protect\subref{fig:cuboids} Cuboids (CAD objects)}
\label{fig:originalShapes}
\end{figure}

For our study, we wished to represent characteristics from these general classes but also allowing for repetition of object types. We wished to measure variation within classes as well as user progress for the same type of object. The chosen objects and their categories are shown in Figure \ref{fig:originalShapes}.

Since at the moment, our focus is on robotics for indoor environments, we did not consider architectural models nor outdoor environments. These, however, do not necessarily fall outside the capabilities of PCP. We therefore chose four types of objects where the categories of \emph{CAD}, \emph{Synthetic}, \emph{Organic}, and \emph{Indoor} would be represented:

\begin{itemize}
\item Indoor environments: Most of the objects we used (except for the CAD category) could be commonly found in an office or a kitchen environment (See Figures \ref{fig:bottles}, \ref{fig:mugs}, \ref{fig:handles} ).
\item Organic shapes: We used two bottles and two mugs. While the bottles had more regular curvilinear parts, the handles on the mugs show more stylized curves and angles (See Figure \ref{fig:mugs}).
\item Synthetic shapes: In addition to the bottles and mugs (which have synthetic qualities), we used a category we called \emph{handle}. This was represented by a shoe-brush and a handle one could find in a drawer. These two are the most different instances within a category. This is due to the fact that the \emph{function} they are meant to be used with is different (See Figure \ref{fig:handles} ).
\item CAD model: We used the \emph{cuboid} class, represented by two objects composed of the boolean operation of three cubes in two different ways (See Figure \ref{fig:cuboids} ).
\end{itemize}

We used the same four types of objects in two separate reconstruction stages. This was done so to measure user progress within the same shape category. In addition to the shape categories, the instances themselves had important characteristics. The bottles had several small details that were below the Kinect's resolution of $2 mm$. The mugs had the added difficulty of representing a different topology (toroidal). The handles, were of different types: with one being the top part of a shoe-brush and the other being a handle similar to the ones used in briefcases. The cuboids, while simple, offered the widest range of construction approaches and were not an object that would be familiar to the subjects. In addition to their shape characteristics, these objects were chosen for the different grasps they might afford.

Since we wanted to have a ground-truth for the dimensions of the objects, we had them 3D-printed from ideal meshes. These 3D-printed objects were then carefully scanned using PCL's Kinfu algorithm to generate the base point clouds that PCP users would work on. In addition to the base point clouds, the mesh output of Kinfu was carefully segmented (to isolate the target shapes) and post-processed to generate a proper watertight model (necessary for extracting some integral properties). These were later used to compare the results of PCP.

In preliminary tests on the point cloud editing capabilities we found that point cloud editing actions did not constitute a major time sink (less than one minute). We therefore pre-segmented the objects from the scene so that subjects could focus on reconstruction routines and not point cloud editing.

For the Solidworks meshes, candidates with experience using the software were asked to first measure, and then design the objects following the same two stages. They, however were allowed to take as much time as necessary to measure the objects. They were also allowed to take as many pictures of the objects as necessary to have a good idea of the shape when they attempted to design the objects.

\paragraph{Metrics:}
During reconstruction, an action log was recorded for all users. In addition, notes were compiled on reported difficulties and user interactions. These were used to obtain number and distribution of actions as well as duration information for every session. This in turn was used to obtain metrics for the effort required to complete each shape.

For the quantitative analysis of shape quality, we used several shape characteristics obtained from the resulting scaffold-generated meshes. We used the work of Brian Mirtich \cite{mirtich1996fast} to compute integral properties from the generated polyhedral masses. These included Center Of Mass (COM),  \textit{surface}, \textit{volume}, and the distribution of point-masses around the volume, as represented by the object's Inertial Tensor (IT). We compared these with those obtained from the original ``ideal'' meshes. The direct error metrics were: 

\begin{itemize}
\item COM Error: $COM_e = d\left( COM_{ideal}, COM_{subject} \right) $
\item Surface Error: $S_e = S_{ideal} - S_{subject}$
\item Volume Error: $V_e = V_{ideal} - V_{subject}$
\item Inertia Tensor Error: $IT_e = \left\| IT_{ideal} - IT_{subject} \right\| $
\end{itemize}

where:

\begin{itemize}
\item[-] $d(\cdot,\cdot)$ is the euclidean distance.
\item[-] $\left\| \cdot \right\|$ is the $L2$ norm.
\end{itemize}

One additional metric is the Hausdorff Distance (HD), which is a popular method of measuring distance between point sets. It is the greatest of all the distances from a point in one set to the closest point in the other set. 
\[ 
H(X,Y)=\max\{\,\sup _{{x\in X}}\inf _{{y\in Y}}d(x,y),\,\sup _{{y\in Y}}\inf _{{x\in X}}d(x,y)\,\}{\mbox{,}}
\]

While useful, this metric is very sensitive to outliers and holes. We therefore used the Mean Hausdorff Distance (meanHD) or $\mu H$ \cite{Curic2014}:

\[ 
\mu H(X,Y)=\max\{\,Mean \left(  \inf _{{y\in Y}}d(x,y) \right) ,\, Mean \left( \inf _{{x\in X}}d(x,y)\,\right)  \}{\mbox{,}}
\]

These measurements can help indicate the size and direction of error. The volume, for example, can be underestimated or overestimated, which is why observing the sign of the results is important. One problem with these measurements is that every object would contain errors at different scales, and therefore would not allow their proper comparison. This is why we obtained a second level of measurements that scaled each metric according to the appropriate characteristic in the ideal meshes. The normalized metrics were:

\begin{itemize}
\item relative COM Error: $rCOM_e =  \left| COM_e \right| /\, diag_{BB}  $
\item relative Surface Error: $rS_e =\left| S_e \right| /\, S_{ideal} $
\item relative Volume Error: $rV_e = \left| V_e \right| /\, V_{ideal}$
\item relative Inertia Tensor Error: $rIT_e =\left| IT_e \right| /\, \lambda_{ideal}^{max} $
\item relative mean Hausdorff Distance: $r\mu H = \left| \mu H \right| /\, diag_{BB}$
\end{itemize}

where:

\begin{itemize}
\item[-] $\left| \cdot\right| $ is the absolute value
\item[-] $diag_{BB}$ is the length of the ideal object's bounding box diagonal.
\item[-] relative Volume Error: $rV_e = \left| V_e \right| /\, V_{ideal}$
\item[-] $\lambda_{ideal}^{max}$ is the maximum eigenvalue in the ideal inertia tensor. This represents the largest wrench that can be applied when the $IT_{ideal}$ is aligned with the world axes.
\end{itemize}

\subsubsection {User Study: PCP Pick-and-Place}

In a way similar to the one used for reconstruction, we devised a protocol that would allow the recording of qualitative and quantitative measurements of the interaction. The objective was to evaluate if the minimal GUI was sufficient to indicate different grasps. In addition, we wished to see what aspects of the task were difficult for humans to deal with and which were easy. To accomplish this, we gathered grasping annotations with PCP and evaluated them using the PR2. In addition, as a comparison interaction scheme, we had subjects physically guide the PR2 arm and gripper to accomplish the same tasks, in a manner similar to \cite{Balasubramanian2014}. In this study it was reported that human guidance was superior to automatic grasping when considering real grasping situations and the larger context of the grasp with respect to the task. For both approaches, we recorded te PR2 end-effector poses throughout the indicated trajectories, from which we were able to extract measurements to compare both approaches. We also recorded video of the live-manipulation so that we could evaluate the quality of the interaction. The protocol for the live-guidance tests is explained in \ref{sec:pr2RigidObjectHandling}.

\paragraph{Experimental Setup (PCP):} 

Participants use a normal keyboard and mouse setup in front of a screen. They first must complete a tutorial that uses videos and asks them to follow along with PCP. Once they have completed this stage, thy are shown each of the tasks and asked to compete them using PCP. No time constraints were imposed on them. 

\paragraph{Protocol (PCP):} 

For handling, subjects were asked to follow a tutorial that would teach them to use PCP. The tutorial lasted between 20 and 30 minutes and was designed to show the features that would allow users to record waypoints for grasping and object \textit{ghost} manipulation. The same way as above, this module could be easily adapted for use in crowdsourcing. 

After the tutorial, subjects were asked to complete manipulation challenges in a virtual setting. The scenario was obtained in the following way: A PR2 robot is placed in front of a table that has some objects placed on its surface. This scenario was captured using PCL's Kinfu and visualized in PCP. Then, subjects were asked to complete a series of manipulation challenges. The challenges were divided into three stages. In the first, they were asked to record a grasp position followed by a waypoint sequence to place an object in a box in one configuration. The second staged asked the same, but with the box placed in a different configuration. In the third, thew were shown a video of specific manipulations (done with the PR2), with their objective being to replicate the sequence as closely as possible.

\begin{figure}
\centering
\subfloat[][]{\includegraphics[width=0.8in]{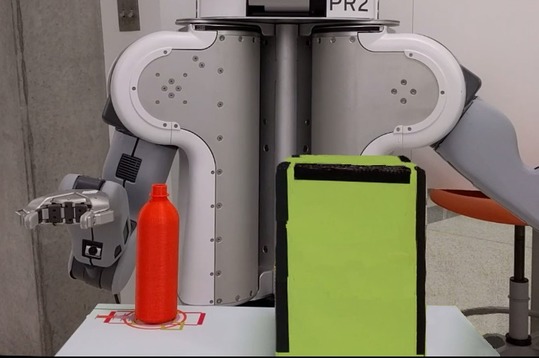}\label{fig:b1rv}}
~
\subfloat[][]{\includegraphics[width=0.8in]{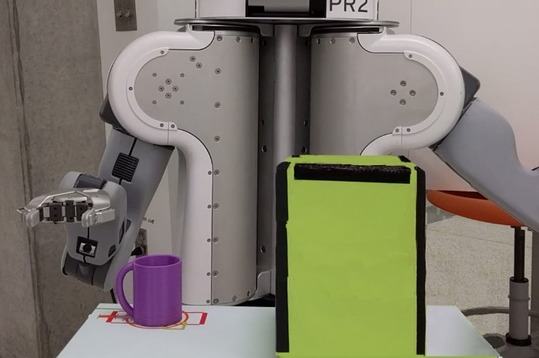}\label{fig:m1rv}}
~
\subfloat[][]{\includegraphics[width=0.8in]{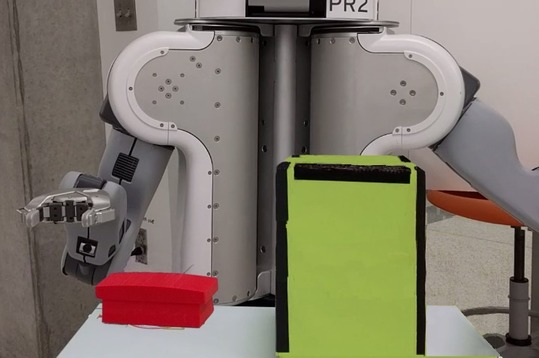}\label{fig:h1rv}}
~
\subfloat[][]{\includegraphics[width=0.8in]{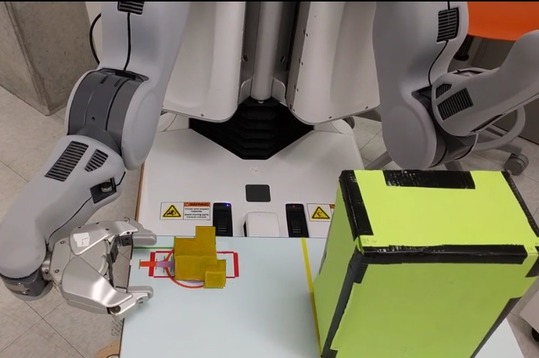}\label{fig:c1rv}}
\caption{Vertical box challenges}
\label{fig:verticals}
\end{figure}

The first and second stages included all eight objects described above, and a box that can be placed with its opening facing forward (vertical box), like a cubby in a bookcase; or facing upward (horizontal box), like a drawer. Figure \ref{fig:verticals} shows the positions of the objects for the vertical challenges.

\begin{figure}
\centering
\subfloat[][]{\includegraphics[width=0.8in]{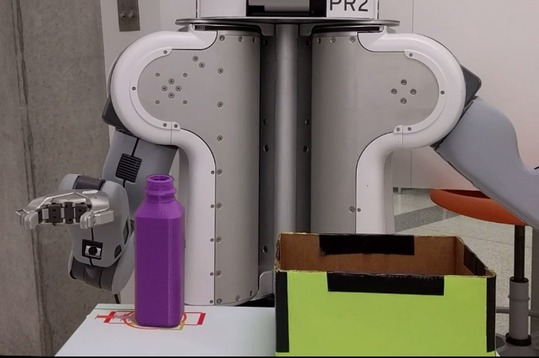}\label{fig:b2rh}}
~
\subfloat[][]{\includegraphics[width=0.8in]{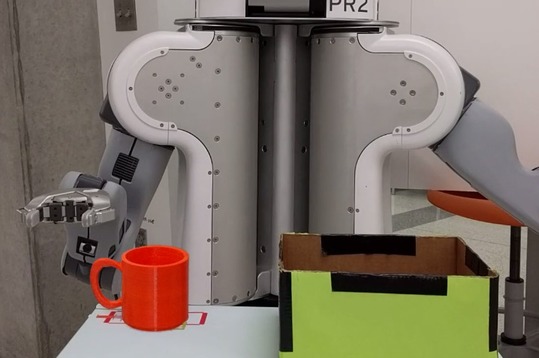}\label{fig:m2rh}}
~
\subfloat[][]{\includegraphics[width=0.8in]{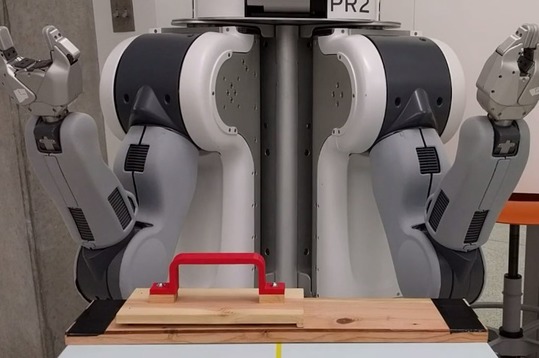}\label{fig:h2rh}}
~
\subfloat[][]{\includegraphics[width=0.8in]{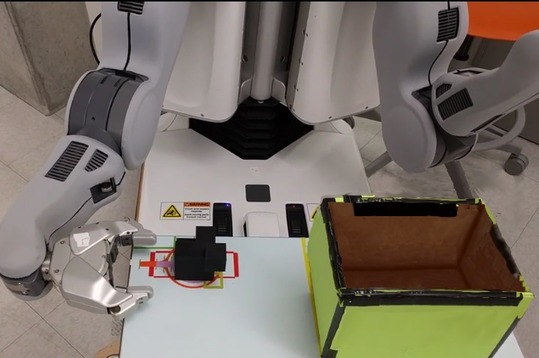}\label{fig:c2rh}}
\caption{Horizontal box challenges}
\label{fig:horizontals}
\end{figure}

Figure \ref{fig:horizontals} shows the positions of the objects for the horizontal challenges. 
 
Stages $1$ and $2$ one had the following challenges:

\begin{itemize}
\item Bottle 1 had to be picked up and placed in the vertical box. See Figure \ref{fig:b1rv}.
\item Mug 1 had to be picked up and placed in the vertical box. See Figure \ref{fig:m1rv}.
\item Handle 1 (brush) had to be picked up used to ``clean'' the near side of the vertical box.See Figure \ref{fig:h1rv}.
\item Cuboid 1 had to be picked up and placed in the vertical box. See Figure \ref{fig:c1rv}.\\
\item Bottle 2 had to be picked up and placed in the horizontal box. See Figure \ref{fig:b2rh}.
\item Mug 2 had to be picked up and placed in the horizontal box. See Figure \ref{fig:m2rh}.
\item Handle 2  had to be grabbed and pulled to rotate a board attached to the table with a hinge. See Figure \ref{fig:h2rh}.
\item Cuboid 2 had to be picked up and placed in the horizontal box. See Figure \ref{fig:c2rh}.
\end{itemize}

The handle challenges were the most specific in terms of functional requirements, while the others just needed to be completed while avoiding major collisions that would cause the challenge to be considered a failure. The reason for the change in box configuration was to see if users would adapt the grasp positions to the new tasks. When assigning the tasks, we always specified the conditions for success:  The grasp had to succeed, the trajectory had to avoid collisions and remain within reach of the robot, and the objects had to be handled in a way that did not conflict with any conditions we attached to the task. We called these \textit{functional objectives} For example, all bottles and mugs were to be handled as if they were carrying water; the brush had to touch the box surface and complete a full sweep of its side; and the rotating board had to rotate along its proper mechanical axis of rotation.

For the third stage, the challenges were the following:
\begin{itemize}
\item Bottle 1 had to be picked up from the top and placed in the horizontal box.
\item Mug 1 had to be picked up from the top and placed in the horizontal box.
\item Handle 2 had to be grabbed up from the top and pulled to rotate a board attached to the table with a hinge.
\item Cuboid 2 had to be picked up from the side and placed in the vertical box.
\end{itemize}

\begin{figure}
\centering
\subfloat[][]{\includegraphics[width=0.8in]{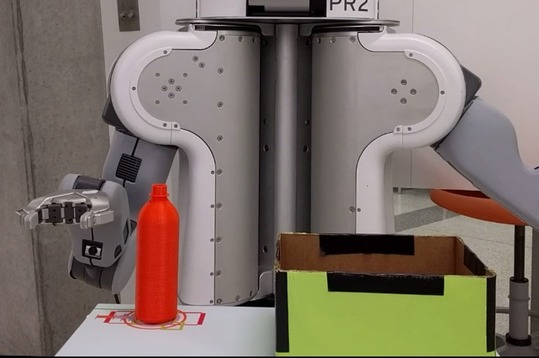}\label{fig:rb1}}
~
\subfloat[][]{\includegraphics[width=0.8in]{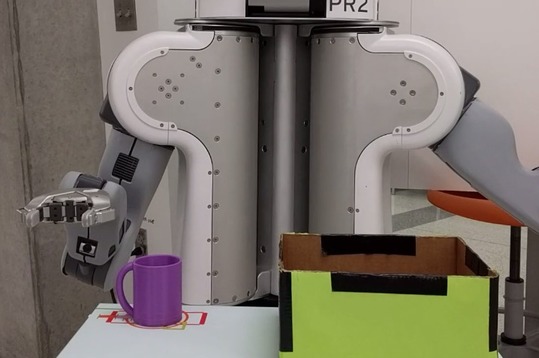}\label{fig:rm1}}
~
\subfloat[][]{\includegraphics[width=0.8in]{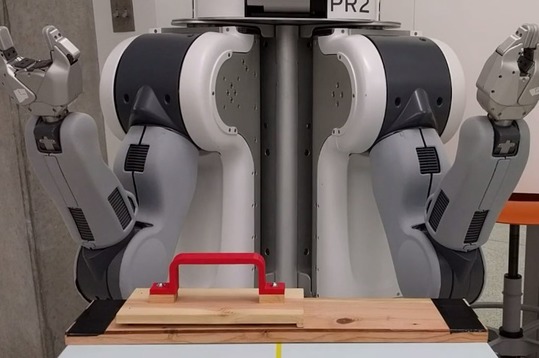}\label{fig:rh2}}
~
\subfloat[][]{\includegraphics[width=0.8in]{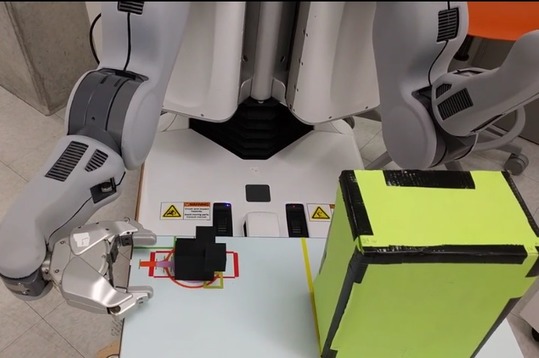}\label{fig:rc2}}
\caption{Replication challenges}
\label{fig:replications}
\end{figure}

These challenges were added to measure the precision of waypoint positioning. These also serve as a way to see what aspects of the interaction are difficult for humans to replicate using a GUI as compared to live manipulation. Figure \ref{fig:replications} shows the setup for these challenges.

\paragraph{Metrics (PCP):} 
We recorded actions and timing in a log during the motion annotation. We also evaluated each grasp and manipulation qualitatively, recording a variety of interaction features that could help characterize the  quality of the manipulation. 

To validate the manipulation annotations, we used the \textit{MoveIt} package in ROS \cite{sucan2013moveit} to follow the set of indicated waypoints. No additional collision avoidance mechanisms were used so that all user-indicated paths with minor or major collisions could be observed and graded. The following are scales used for classifying the interaction quality.

\subparagraph{Grasping:} 

We divided the qualitative analysis of the grasps into \textit{pure grasp quality} and \textit{functional quality}. Pure grasp quality evaluates the grasp by itself, and whether or not it can hold the object independent of the handling task. The scaling for pure grasp quality is:

\begin{enumerate}
\item[(1)] Miss: the object is missed completely by the PR2.
\item[(2)] Slip: the object is nudged or loosely grasped and dropped at some point.
\item[(3)] Shift: the grasp has minor flaws but holds.
\item[(4)] Good: the object is firmly grasped.
\end{enumerate}

Functional quality looks at whether or not the grasp would allow the proper completion of the task independent of the path used to do so. As an example think of a mug held by the top. If the objective is to pour, then the grasp would receive a good pure grasp quality but a flawed or impossible functional quality, since the top grasp would make the pouring difficult or impossible. For this evaluation, the scales are the following:

\begin{enumerate}
\item[(1)] Impossible: the grasp would prevent the completion of the task.
\item[(2)] Flawed: the grasp has minor flaws but can allow the completion of the task.
\item[(3)] Good: the grasped is adequate for the task.
\end{enumerate}

For the quantitative analysis of grasp quality, we used the metrics provided by GraspIt: the volume of the Grasp Wrench Space and the radius of the largest inscribed sphere, also called the epsilon-distance. GraspIt has the capacity to verify grasps under dynamic settings, allowing the objects and grippers to shift into stable grasps.

\emph{Force-Closure} is a grasp that can resist disturbance forces from any direction, given sufficiently large contact forces. We considered grasps with force closure under dynamic settings as successful grasps.

\subparagraph{Handling:}

For the Pick-and-Place manipulation as a whole, we looked at three different handling characteristics: collisions, robot configuration problems, and object handling flaws. 

\subparagraph{Collisions:} 

\begin{enumerate}
\item[(1)] Major: the gripper hits the table, object, or box in a way that makes the challenge fail.
\item[(2)] Minor: the gripper nudges the object or table mildly, in a way that does not preclude success.
\item[(3)] Free: no collisions.
\end{enumerate}

\subparagraph{Robot configuration problems:} 

\begin{enumerate}
\item[(1)] Fail: for most or crucial parts of the path, the robot would hit itself or attempts to move beyond its reach.
\item[(2)] Partial: at a single uncritical part of the path, the robot would hit itself or attempts to move beyond its reach.
\item[(3)] Good: no problems with self collisions or reach.
\end{enumerate}

\subparagraph{Object handling flaws:}

\begin{enumerate}
\item[(1)] Bad: for most or crucial parts of the path, handling of the object invalidates the functional objectives.
\item[(2)] Fair: at a single uncritical part of the path, object is mildly mishandled.
\item[(3)] Good: no problems with object handling.
\end{enumerate}

The combination of the above three requirements would determine the result of the overall object manipulation challenge (discounting grasping). Path quality had the following scale:

\subparagraph{Path quality:}

\begin{enumerate}
\item[(1)] Impossible: the task cannot be completed with the specified path.
\item[(2)] Flawed: parts of the annotation cause the path to be executed imperfectly.
\item[(3)] Good: path is adequate for the task.
\end{enumerate}

Finally, the whole task, including grasping and path handling were combined to evaluate the quality of the challenge as a whole.

\subparagraph{Challenge quality:}

\begin{enumerate}
\item[(1)] Fail: the challenge cannot be completed with the specified path or grasp.
\item[(2)] Partial: parts of the annotation cause the challenge to be executed imperfectly.
\item[(3)] Good: the path and grasp are adequate for completing the challenge.
\item[(4)] Perfect: The challenge is completed with no flaws.
\end{enumerate}

In addition to these manipulation characteristics we extracted the total distance of the chosen manipulation. Also, for the replication challenges, we developed a path comparison metric that we called \textbf{ribbon area}. 

\begin{figure}
\centering
\subfloat[][]{\includegraphics[width=1.5in]{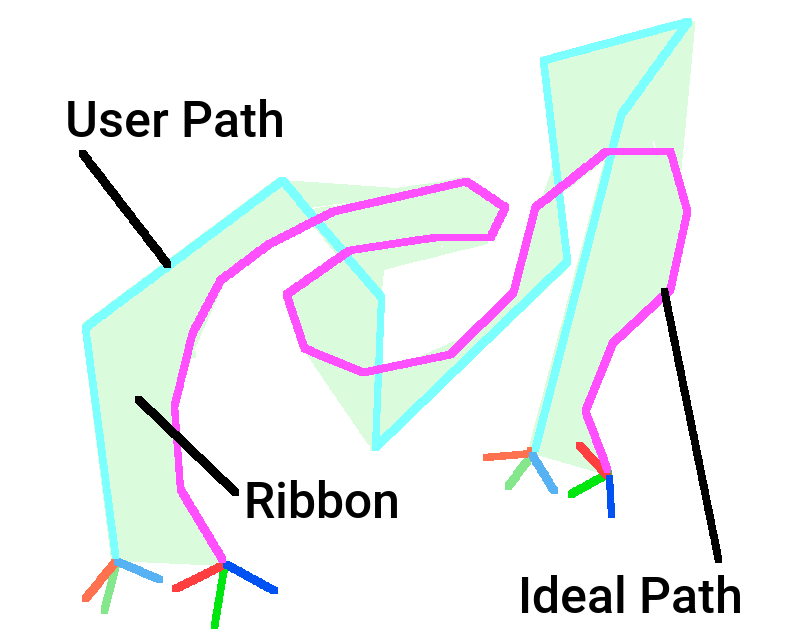}\label{fig:IdealRibbon}}
\subfloat[][]{\includegraphics[width=1.5in]{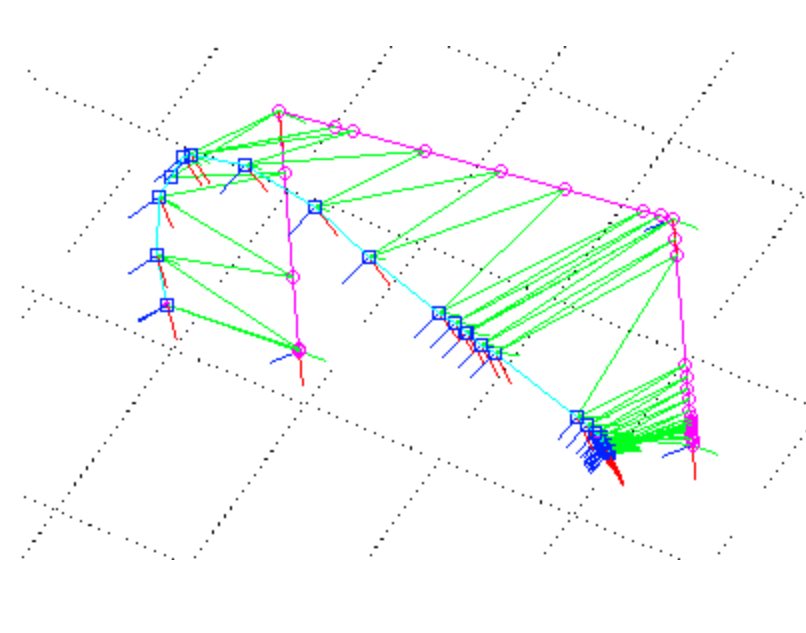}\label{fig:matlabRibbon}}
\caption{Ribbon area concept  \protect\subref{fig:IdealRibbon} and implementation  \protect\subref{fig:matlabRibbon}}
\label{fig:ribbon}
\end{figure}

\subparagraph{The ribbon area:} path error was defined as the line integral of the distance function between the two paths. In practice, and since we are using discrete waypoints, we simplified this to simply add the area generated between subsequent pairs of points at equivalent positions along the paths. his area becomes zero for identical paths, and grows as the distance between equivalent points increases. Figure \ref{fig:IdealRibbon} shows the concept;  Figure \ref{fig:matlabRibbon} shows the implementation using Matlab;

\subsubsection {User Study: PR2 Rigid Object Handling} \label{sec:pr2RigidObjectHandling}

\paragraph{Experimental Setup (PR2):} 

\begin{figure}[!htbp]
\centering
\includegraphics[width=2in]{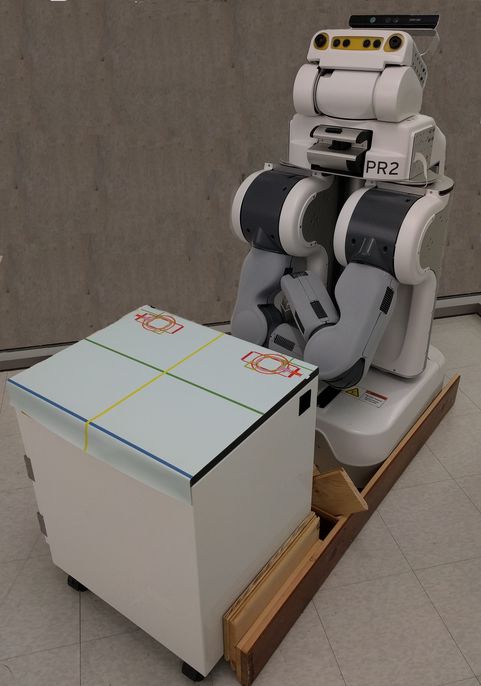}
\caption{PR2 with harness and table with shape projections}
\label{fig:PR2setup}
\end{figure}

Figure \ref{fig:PR2setup} shows the PR2 handling setup, and Figures \ref{fig:verticals}, \ref{fig:horizontals}, and \ref{fig:replications}, are the (same) setups for the PR2 challenges. The PR2 was always positioned in a pre-defined initial configuration to maximize its reach. The table in front of it was fixed in relation to the robot using a harness to prevent the variation in the relations between parts. Objects were placed in the same locations, guided by their projected positions on the worktable. 

\paragraph{Protocol (PR2):} 

For handling, subjects were shown how to move one arm of the PR2 robot (the same as their dominant hand). The tutorial lasted between 5 and 10 minutes and was designed to familiarize the subject with the ways in which the robot could move. While the PR2 robot has a similar general structure to that of a person, the PR2 arm joints have slightly different motion capabilities that human arms i.e. some joints, like the wrist or the shoulder are composed of two independent revolute joints. 

Subjects were instructed to complete the same 12 challenges as the ones described above. The sequence of actions was the following:

\begin{enumerate}
\item The subject had to move the arm from an initial position to a grasp position of their choice. 
\item When ready, they would instruct the PR2 operator to close the gripper.
\item The subject would then move the arm and the held object in a path that would complete a challenge.
\item When done, the subject would ask the PR2 operator to open the gripper.
\item After releasing, they would return the arm to the initial position.
\end{enumerate}

\paragraph{Metrics (PR2):} 

We used the same quality scoring system as the one specified above. Since we saved the end-effector path as sequences of poses, we were able to convert the physical motion to a sequence of waypoints that could be directly compared to the PCP manipulation experiments.

\subsection{Data Analysis}

\subsubsection{Reconstruction Statistics}

The experimental design is for \textit{independent measures} with many groups. As will be explained in the Results section, the error values did not have a normal distribution. We therefore used the non-parametric Kruskal-Wallis test \cite{hollander2013nonparametric} to verify if the population distributions for all approaches were identical. If a difference was found, pairwise Mann-Whitney-Wilcoxon Test \cite{bauer1972constructing} with Bonferroni correction \cite{dunn1958estimation} was employed to see if pairs of approaches could be considered statistically identical or not.

For the case of analyzing interactions between variables, we fitted a linear model to complete a regression analysis between pairs of variables. We used the Shapiro-Wilk normality test to verify that the regression residuals are well behaved and no important variables are missing.

\subsubsection{Handling Statistics}

This analysis applied to both the PCP and PR2 handling modules. The experimental design is for \textit{independent measures} with many groups. Here, we used frequencies to record successes over failures for each challenge. We therefore used the Chi-squared test of independence to verify if the population distributions for all approaches were identical. If a difference was found, post-hoc pairwise Chi-squared tests with Bonferroni correction were performed. 

For the grasp objective metrics, we used the non-parametric Kruskal-Wallis test to verify if the population distributions for all approaches were identical. If a difference was found, the pairwise Mann-Whitney-Wilcoxon Test with Bonferroni correction was employed to see if pairs of approaches could be considered statistically identical or not.

For the case of analyzing interactions between variables, we fitted a linear model to complete a regression analysis between pairs of variables. We used the Shapiro-Wilk normality test to verify that the regression residuals are well behaved and no important variables are missing.

\section{Results \label{sec:results} }

This section details the results of the qualitative and quantitative analysis of four tasks: reconstruction, grasping, and Pick-and-Place.

\subsection{Reconstruction Results}

\subsubsection{Competing Methods}
\begin{itemize}
\item PCP : These are novice users of PCP.
\item PCP Expert : This is an expert user of PCP.
\item CAD : These are experienced Solidworks users.
\item Kinfu : This is PCL's version of the Kinect Fusion automatic reconstruction algorithm.
\end{itemize}

For PCP, the total number of participants to complete the minimum number of shapes ($3$) was $N_{PCP}^U = 14$. A total of $N_{PCP}^R = 94$ novice user shapes were created, and since we used relative error metrics, we were able to integrates measurements from all the different shapes.
For cad, $N_{CAD}^U = 5$ subjects provided shapes. They were selected from students that had completed a CAD modeling class using Solidworks, and ranged in experience from $30$ to $100+$ hours.  A total of $N_{CAD}^R = 40$ CAD user shapes were created. Again, the use of relative error metrics allowed us to integrates measurements from all the different shapes.
The last shapes we evaluated were the \textit{mean shapes} for each object. We obtained a total of $N_\mu^R=8$ \textit{mean shapes}, each of which required an average of $N_{i}^{U} = 11.75$ novice meshes (since not all shapes were completed by all subjects).

\subsubsection{Qualitative Analysis of Shapes}
\begin{figure}
\centering
\subfloat[][]{\includegraphics[width=1.5in]{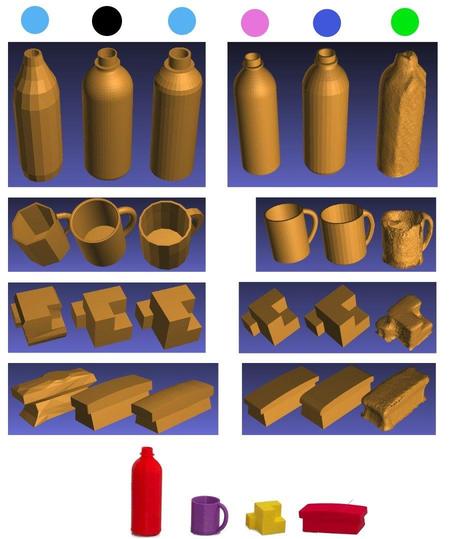}\label{fig:shapesS1}}
\subfloat[][]{\includegraphics[width=1.5in]{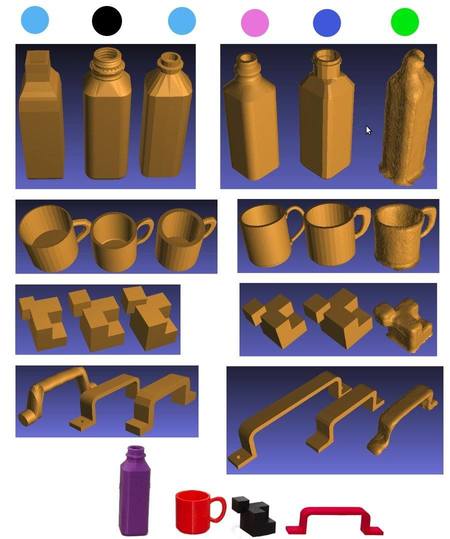}\label{fig:shapesS2}}
\caption{ Examples of reconstructed shapes for the first \ref{fig:shapesS1} and second  \ref{fig:shapesS2} stages.  a black dot is the original object; light blue is for good and bad novice users in PCP; the dark blue dot is for the PCP expert shapes; the pink dot is for the Solidworks users; and the green one is the Kinfu reconstruction}
\label{fig:reconstructions}
\end{figure}

While subjective, visual inspection of the constructed shapes revealed several interesting features. Figure \ref{fig:reconstructions} shows examples of reconstructed shapes for all approaches. Figure  \ref{fig:shapesS1} shows the result of stage $1$, and \ref{fig:shapesS2} shows the result of stage $2$. In the images, the column under the black dot is the original object; the light blue column represents examples of good and bad novice user reconstructions; the dark blue dot is for the PCP expert shapes; the pink dot is for the Solidworks users; and the green one is the Kinfu reconstruction.

\paragraph{Kinfu:} First, the obvious problem with Kinfu has to do with the aforementioned issues with thin surfaces, occlusions, and sensitivity to scan smoothness. The Kinfu scans tends to underestimate the  \textit{external} volume of shapes, which, in turn, affects the modeling on PCP, which uses the Kinfu-generated pointclouds as the basis of the scaffold placement. On the other hand, Kinfu does not have the ability to record internal structures or hard-to-see cavities, which means that it would consider those areas as ``occupied'', and therefore as extra positive volume. As a result, shapes with cavities like the bottles and mugs are overestimated in terms of overall volume, while shapes without cavities are underestimated in volume.

\paragraph{Solidworks:} These were expected to be the most accurate shapes, but, even though they are in general, one bad measurement can result in large errors in the resulting shape. 

\paragraph{PCP novices: } The initial shapes vary greatly in quality for the first set of four, but become less varied for the final four shapes. The greatest errors occur for the initial shapes, like erroneous hole specification, or problems with the generation of some object's part. 

Compare the results obtained in PCP \ref{fig:reconstructions} with those from crowdsourcing reconstructions obtained from \cite{Sorokin2010}.

\begin{figure}[!htbp]
\centering
\includegraphics[width=3in]{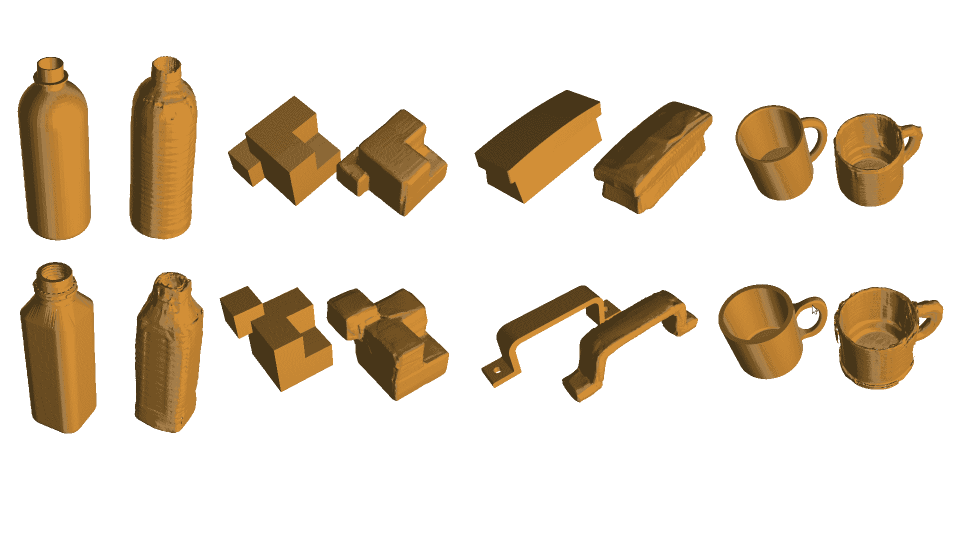}
\caption{Poisson reconstruction o of the merged novice PCP shapes }
\label{fig:mixedPCPshapes}
\end{figure}

\paragraph{PCP mean shapes:} The quality of the resulting shapes was quite surprising. The shapes can be seen in Figure \ref{fig:mixedPCPshapes} next to the original objects. While the novice pcp shapes contained outliers with large errors, the mean shapes contain more similar appearances than the individual novice reconstructions. Locations where multiple individual shapes ``agree'' on shape are reinforced and are therefore maintained in the resulting mean shape, causing outliers to be dismissed.

\subsubsection{Quantitative Analysis of Shapes}

The following are the results from comparing the shape measurements. As mentioned before, we used the relative metrics to be able to compare the different shapes. These do not follow normal distributions, and therefore had to be analyzed using a non-parametric test. We used the Kruskal-Wallis test to determine if the population distributions were identical. If they were not, a pairwise Wilcox test with Bonfernoni correction was used to determine which pairs of treatments were statistically different.

In the following, \textit{CAD} represents the results from the Solidworks shapes; \textit{Kinfu} for the automatic reconstruction; \textit{PCP} for the novice PCP users; $\mathit{PCP_E}$ is for the PCP expert; and $\mathit{\mu PCP_E}$ for the PCP mean shapes. The first thing we evaluated was the difference in modeling times for the Human-In-the-Loop approaches: CAD, PCP, and PCP expert.

\paragraph{Full Modeling Time: }
We found a significant difference in modeling times ($\chi^2_4 = 35.4$, $p<0.0001$).
Table \ref{tab:RecTime} shows the result of applying the pairwise Wilcox test with Bonferroni correction .

\begin{table} [!htbp]
\centering
\begin{tabular}{|c|c|c|}
\hline         & CAD     & PCP    \\
\hline PCP     & 1.000   & -      \\     
\hline $PCP_E$ & <0.05  & <0.01 \\ 
\hline 
\end{tabular} 
  \caption{relative volume error per approach} \label{tab:RecTime}
\end{table}

\begin{figure}[!htbp]
\centering
\includegraphics[width=3in]{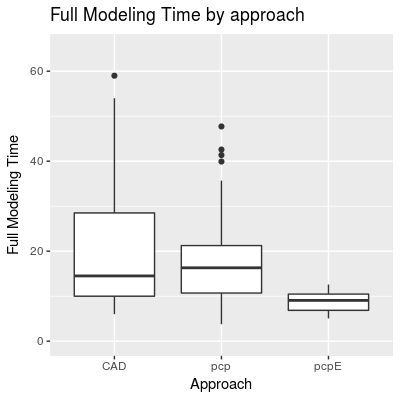}
\caption{Modeling duration for different approaches}
\label{fig:RecTime}
\end{figure}

As can be seen from the table and Figure \ref{fig:RecTime}, PCP and CAD users took about the same time to make the shapes, while the PCP expert took less time. This result was somewhat influenced by the modeling protocol, where we indicated time limits for each modeling session.

\begin{figure}[!htbp]
\centering
\includegraphics[width=3in]{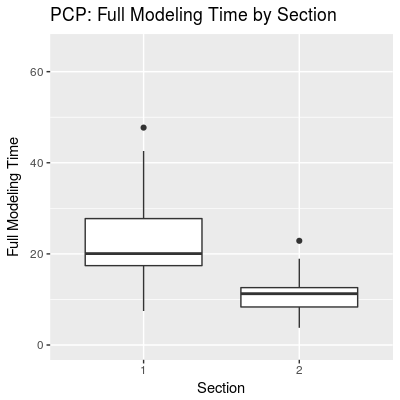}
\caption{PCP only: Modeling duration for each modeling section}
\label{fig:RecTimeSections}
\end{figure}

As can be seen in Figure \ref{fig:RecTimeSections}, modeling times were more variable in the first section, which points to a rapid homogenization after only four modeling tasks. 

\paragraph{Error Metrics:} 
The next section show the differences for each error metric.

\paragraph{Center of Mass:} 
We found a significant difference in COM errors ($\chi^2_4 = 40$, $p<0.0001$). 
Table \ref{tab:COMEp} shows the result of applying the pairwise Wilcox test with Bonferroni correction .
\begin{table} [!htbp]
\centering
\begin{tabular}{|c|c|c|c|c|}
\hline         & CAD     & kinfu & PCP   & $PCP_E$     \\
\hline Kinfu   & 0.240   & -     & -     & -    \\     
\hline PCP     & <0.0001 & 1.000 & -     & -    \\     
\hline $PCP_E$  & 1.000  & 1.000 & 0.078 & -    \\ 
\hline $\mu$PCP & 1.000  & 1.000 & 0.216 & 1.000\\ 
\hline 
\end{tabular} 
  \caption{relative volume error per approach} \label{tab:COMEp}
\end{table}

\begin{figure}[!htbp]
\centering
\includegraphics[width=3in]{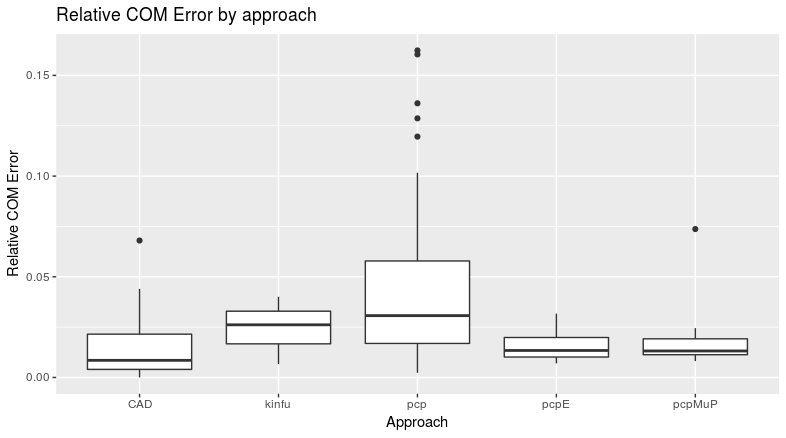}
\caption{relative Center Of Mass error for different approaches}
\label{fig:COMEp}
\end{figure}

This, together with the box plots shown in Figure \ref{fig:COMEp}, indicates that the PCP expert and $\mu$PCP had a similar error than CAD,and while Kinfu and PCP novices fared a bit worse, they had small and similar error sizes among themselves.

\paragraph{Surface:} 
We found a significant difference in surface errors ($\chi^2_4 = 20.3$, $p<0.001$).
Table \ref{tab:surfEp} shows the result of applying the pairwise Wilcox test with Bonferroni correction.

\begin{table} [!htbp]
\centering
\begin{tabular}{|c|c|c|c|c|}
\hline          & CAD     & kinfu & PCP   & $PCP_E$ \\
\hline Kinfu    & <0.05   & -     & -     & -    \\     
\hline PCP      & <0.05   & 0.231 & -     & -    \\     
\hline $PCP_E$  & 1.000   & <0.05 & 0.418 & -    \\ 
\hline $\mu$PCP & 0.888   & 0.148 & 1.000 & 1.000\\ 
\hline 
\end{tabular} 
  \caption{relative volume error per approach} \label{tab:surfEp}
\end{table}

\begin{figure}[!htbp]
\centering
\includegraphics[width=3in]{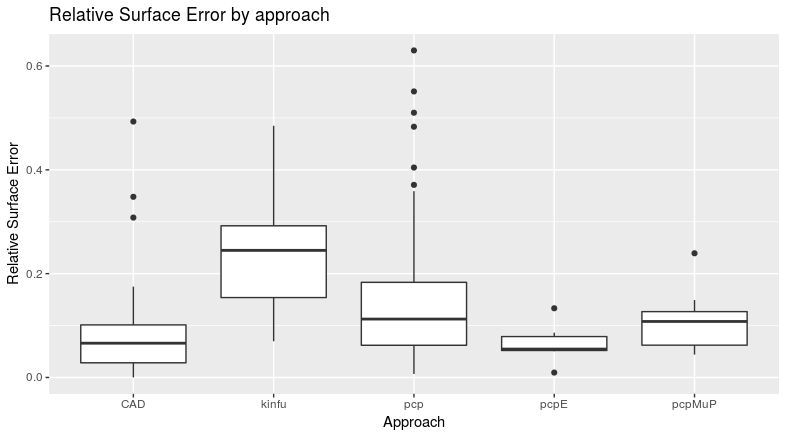}
\caption{relative surface error for different approaches}
\label{fig:surfEp}
\end{figure}

This, together with the box plots shown in Figure \ref{fig:surfEp}, again indicates that the PCP expert and $\mu$PCP had a similar error than CAD. Kinfu and PCP novices also had small and similar error sizes.

\paragraph{Volume:} 
We found a significant difference in volume errors ($\chi^2_4 = 16.7$, $p<0.01$).
Table \ref{tab:volEp} shows the result of applying the pairwise Wilcox test with Bonferroni correction .
\begin{table} [!htbp]
\centering
\begin{tabular}{|c|c|c|c|c|}
\hline          & CAD     & kinfu & PCP   & $PCP_E$ \\
\hline Kinfu    & 0.1050  & -      & -      & -    \\     
\hline PCP      & <0.01  & 1.0000 & -      & -    \\     
\hline $PCP_E$  & 1.0000  & 0.3792 & 0.7509 & -    \\ 
\hline $\mu$PCP & 1.0000  & 1.0000 & 1.0000 & 1.000\\ 
\hline 
\end{tabular} 
  \caption{relative volume error per approach} \label{tab:volEp}
\end{table}

\begin{figure}[!htbp]
\centering
\includegraphics[width=3in]{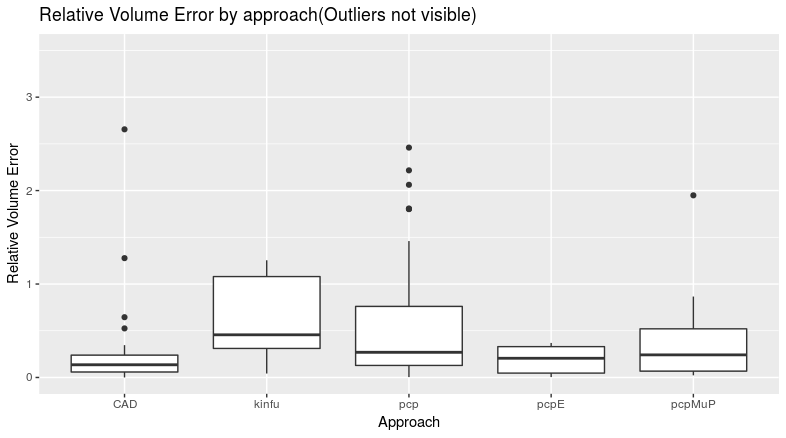}
\caption{relative volume error for different approaches}
\label{fig:volEp}
\end{figure}

This, together with the box plots shown in Figure \ref{fig:volEp}, again indicates that the PCP expert and $\mu$PCP had a similar error than CAD. Kinfu and PCP novices also did a bit worse and had small and similar error sizes.

\paragraph{Inertia Tensor:}
We found no significant difference in IT errors.

\begin{figure}[!htbp]
\centering
\includegraphics[width=3in]{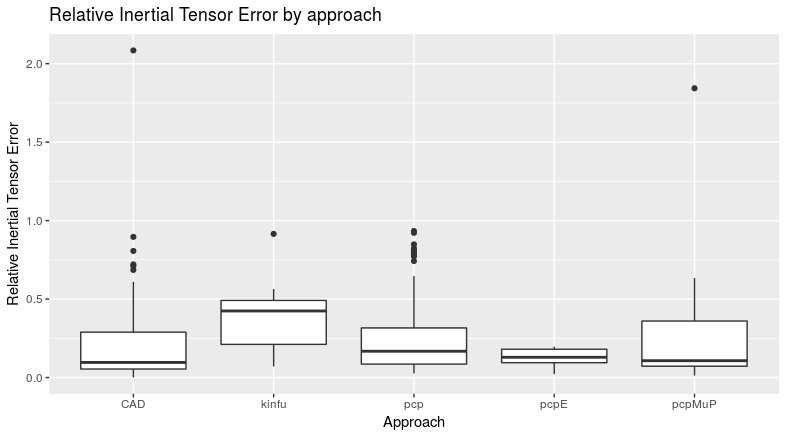}
\caption{relative inertia tensor error for different approaches}
\label{fig:L2Ep}
\end{figure}

The relative Inertia tensor errors shown in Figure \ref{fig:L2Ep} show that the error distributions were similar. One thing to note is that the smallest variation in error was for the PCP expert, which indicates consistency. The variabilities for novice PCP users and CAD users were similar.

% % % % % % % % % % % % % % % % % % % % % 
% % % % % % % % % % % % % % % % % % % % % 
\paragraph{Hausdorff Distance:} 
We found a significant difference in meanHD errors ($\chi^2_4 = 35.5$, $p<0.0001$).
Table \ref{tab:uHDEp} shows the result of applying the pairwise Wilcox test with Bonferroni correction .
\begin{table} [!htbp]
\centering
\begin{tabular}{|c|c|c|c|c|}
\hline          & CAD     & kinfu & PCP   & $PCP_E$ \\
\hline Kinfu    & <0.05  & -     & -      & -    \\     
\hline PCP      & <0.0001 & 1.000 & -      & -    \\     
\hline $PCP_E$  & 1.0000  & <0.05 & <0.05 & -    \\ 
\hline $\mu$PCP & 1.0000  & 0.006 & 0.0731 & 1.000\\ 
\hline 
\end{tabular} 
  \caption{relative volume error per approach} \label{tab:uHDEp}
\end{table}

\begin{figure}[!htbp]
\centering
\includegraphics[width=3in]{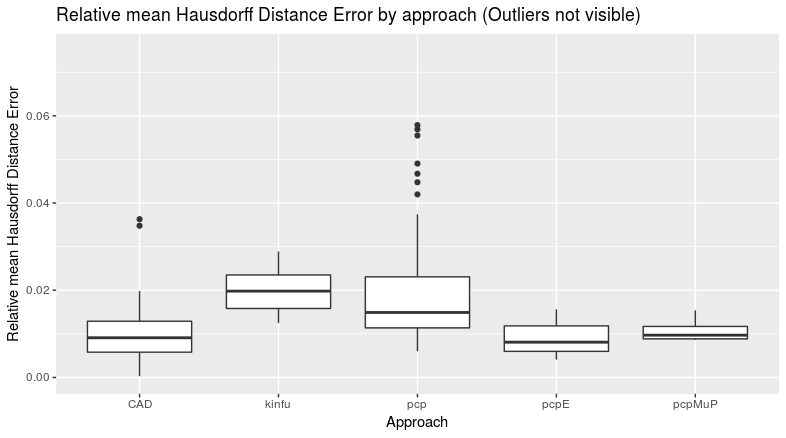}
\caption{relative Hausdorff distance for different approaches}
\label{fig:uHDEp}
\end{figure}

The above table and the box plots from Figure \ref{fig:uHDEp} show that the error distributions for CAD, PCP users, and $\mu$PCP were similar. Kinfu and PCP novices also did a bit worse and had small and similar error sizes.

Interesting general observations to note are that novice PCP users consistently matched or outperformed the automatic method in terms of quality, while the PCP expert proved consistently superior. The PCP expert as well as $\mu$PCP attained a level of precision as good as the CAD users, with the PCP expert times being much shorter than those of CAD. This highlights the surprising and promising approach of merging reconstructions to obtain better prototype objects. The above results hint at the possibility of a rapid increase in precision for new users.

\paragraph{ Efficiency:} 
Since we have duration information and precision errors, we devised a modeling efficiency measurements based on the Hausdorff Distance error metric, which we defined as:

\[ 
M^E_i = \frac{1 - r\mu HD}{duration}
\]

where $pE_i$ represents each of the relative error ratios for COM, surface, etc.

\subparagraph*{Efficiency with respect to the Hausdorff Distance error metric:}

\begin{table} [!htbp]
\centering
\begin{tabular}{|c|c|c|}
\hline         & CAD     & PCP    \\
\hline PCP     & 1.000  & -      \\     
\hline $PCP_E$ & <0.05  & <0.01 \\  
\hline 
\end{tabular} 
  \caption{Efficiency with respect to Hausdorff error per approach} \label{tab:eff_muHDEp}
\end{table}

\begin{figure}[!htbp]
\centering
\includegraphics[width=3in]{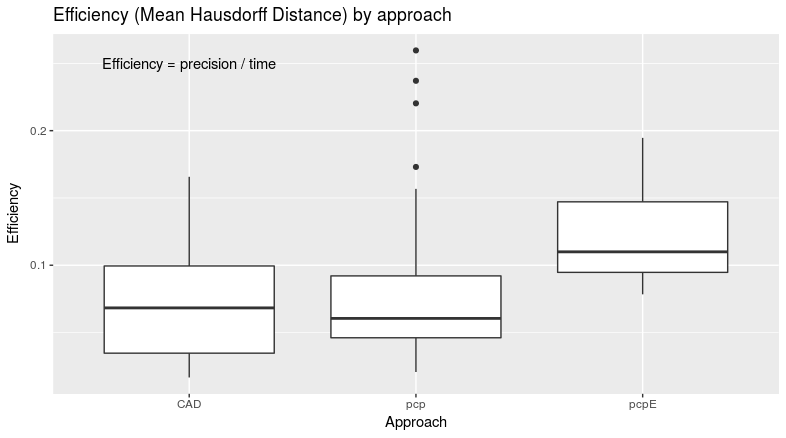}
\caption{Efficiency (relative HD error Metric) for each modeling approach}
\label{fig:eff_muHDEp}
\end{figure}

The above table and Figure \ref{fig:eff_muHDEp} show that novice PCP and CAD users obtain similar efficiency, while the PCP expert surpass both. 

\paragraph{ Learning Effort:} 
The above results show that annotation through PCP has a lot of promise in terms of efficiency and precision. Another important factor is the effort put into learning the software. While no CAD user had less than a minimum of $30$ hours of training(self reported), PCP users had only a 2 hour tutorial. Given the similar accuracy results, this hints at a low learning curve for PCP.

\begin{figure}[!htbp]
\centering
\includegraphics[width=3in]{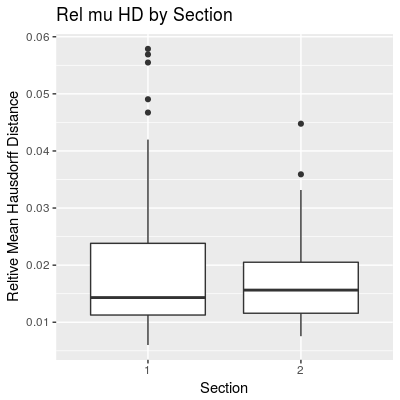}
\caption{PCP only: relative HD error for each section}
\label{fig:PCPqual_muHDEp}
\end{figure}

\paragraph{Quality by section: } 
For novice PCP only, we evaluated the overall shape quality, as indicated by the Hausdorff Distance for each of the two stages. The box plots in Figure \ref{fig:PCPqual_muHDEp} reveals that the errors had the same distribution (no significant difference) despite the accelerated time requirements for the second section. This supports the idea of a low learning curve and that users may attain good precision quickly.

\paragraph{Measured interactions: } In addition to shape characteristics, we analyzed possible interactions between these and the user's spatial reasoning score (SR). 
No correlation was found between SR and COM or Surface errors. However, we found medium to large effects for Volume, IT, Hausdorff Distance, and modeling duration. These confirm the assumption that subjects with higher spatial reasoning abilities obtain better results due to their increased understanding of shape in 3D.

\begin{figure}[!htbp]
\centering
\includegraphics[width=3in]{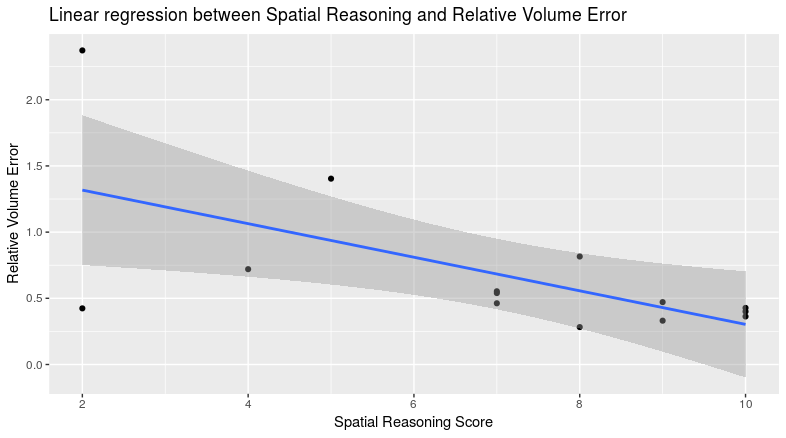}
\caption{PCP only: variation in relative volume error by SR}
\label{fig:PCPreg_volEp}
\end{figure}

\subparagraph*{Relative Volume error vs SR:}
Figure \ref{fig:PCPreg_volEp} shows the results of the linear regression where we found a medium effect size (SQ explains $\approx 39\%$ of variation in relative volume error), $R^2=  0.3898$,	$F_1,12 = 7.665$,  $p<0.05 $.

\begin{figure}[!htbp]
\centering
\includegraphics[width=3in]{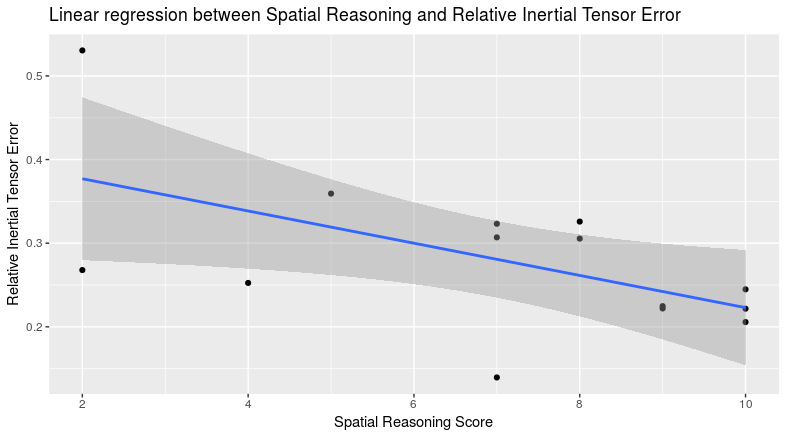}
\caption{PCP only: variation in relative IT error by SR}
\label{fig:PCPreg_L2Ep}
\end{figure}

\subparagraph*{Relative IT error  vs SR:}
Figure \ref{fig:PCPreg_L2Ep} shows the results of the linear regression where we found a small-to-medium effect size (SQ explains $\approx 33\%$ of variation in relative Inertia Tensor error), $R^2=  0.3325$,	$F_1,12 = 5.977$,  $p<0.05 $.

\begin{figure}[!htbp]
\centering
\includegraphics[width=3in]{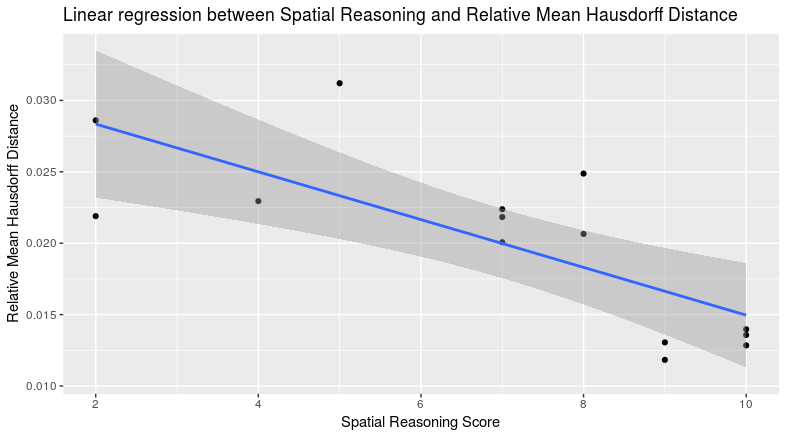}
\caption{PCP only: variation in relative IT error by SR}
\label{fig:PCPreg_muHDEp}
\end{figure}

\subparagraph*{Relative HD error  vs SR:}
Figure \ref{fig:PCPreg_muHDEp} shows the results of the linear regression where we found a large effect size (SQ explains $\approx 57\%$ of variation in relative Hausdorff Distance error), $R^2=  0.5725$,	$F_1,12 = 16.07$,  $p<0.01$. 

\begin{figure}[!htbp]
\centering
\includegraphics[width=3in]{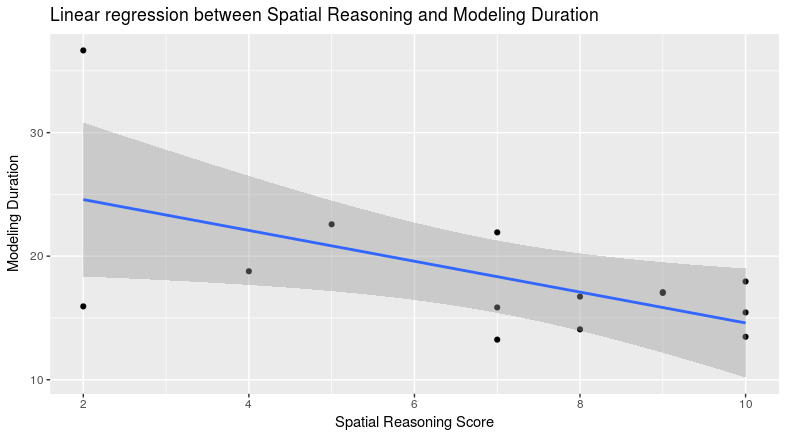}
\caption{PCP only: variation in duration by SR}
\label{fig:PCPreg_dur}
\end{figure}

\subparagraph*{Modeling Duration vs SR:}
Figure \ref{fig:PCPreg_dur} shows the results of the linear regression where we found a small-to-medium effect size (SQ explains $\approx 34\%$ of variation in relative Hausdorff Distance error), $R^2=  0.3382$,	$F_1,12 = 6.132$,  $p<0.05$. 

\subparagraph*{Number of Modeling Actions vs SR:}
No correlation was found between SR and the number of actions a user employed for modeling.

\subsection{Grasping Results}

The approaches considered were PCP, GraspIt!, and manual placement using the PR2 robot. For PCP and the PR2, novice users were compared to the expert user(s). For PCP, the total number of participants to complete the minimum number of grasping tasks ($3$) was $N_{PCP}^U = 16$. A total of $N_{PCP}^G = 183$ novice user grasping poses were recorded and compared. For GraspIt!, the Eigengrasp automatic grasping algorithm was used to generate candidate grasps, where we kept the top $10$ scoring grasp hypotheses. Since we had $12$ grasping challenges, these gave us $N_{GraspIt}^G = 120$ automatic grasp poses to compare. For the PR2, we had  $N_{PR2}^U = 18$ participants, who generated $N_{PR2}^G = 214$ grasps (not all users finished all tasks). 
 
\subsubsection{Qualitative Grasp Analysis}

The PCP grasp suggestions were evaluated in the PR2 by using MoveIt to place the end-effector in the indicated position, closing the gripper, and then proceeding with the rest of the manipulation. In the case of GraspIt, the $10$ top grasp suggestions obtained from Eigengrasp were evaluated using the GWS volume and eps-dist, and were marked as \emph{Good} if the grasp had force-closure (any valid GWS volume), and \emph{Slip} otherwise.

\begin{figure}[!htbp]
\centering
\includegraphics[width=3in]{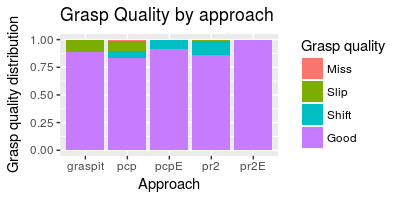}
\caption{Grasp quality by approach}
\label{fig:GC_byApp}
\end{figure}

\paragraph{Grasp Quality Distribution:} Figure \ref{fig:GC_byApp} shows the distribution of observed events during a grasp attempt for each of the grasping approaches.  

\paragraph{Grasp Success Rate:} 
 
\begin{figure}[!htbp]
\centering
\includegraphics[width=3in]{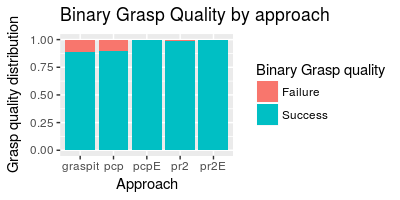}
\caption{Grasp Success rate by approach}
\label{fig:Gsuc_byApp}
\end{figure}

Figure \ref{fig:Gsuc_byApp} shows the distribution of successes vs failures for each of the evaluated approaches. In these, \textit{Good} or \textit{Shift} are marked as a \textit{Success}, and as a \textit{Failure} otherwise.
As can be seen, the majority of actions in every one of the methods result in a grasp that holds the object for a future handling task. 
 
We found a significant difference in grasp success rates ($\chi^2_4 = 20.3$, $p<0.001$).
Table \ref{tab:Gsuc_byApp} shows the result of applying the post-hoc pairwise Chi-squared tests with Bonferroni correction\footnotemark{} . 

\begin{table} [!htbp]
\centering
\begin{tabular}{|c|c|c|c|c|}
\hline         & GRASPIT   & PCP & $PCP_E$ & PR2     \\
\hline PCP     & 1.000     & -       & -     & -    \\     
\hline $PCP_E$ & 1.000     & 1.000    & -     & -    \\     
\hline PR2     & <0.01     & <0.01 & 1     & -    \\ 
\hline $PR2_E$ & 1.000     & 1.000    & -\textsuperscript{1}   & 1.000\\ 
\hline 
\end{tabular} 
  \caption{Grasp Success rate per approach} \label{tab:Gsuc_byApp}
\end{table}

\footnotetext{For some comparisons there was not enough data}

What can be seen from the table \ref{tab:Gsuc_byApp}, and Figures \ref{fig:GC_byApp}, and \ref{fig:Gsuc_byApp}, is that  PCP users perform as well as GraspIt in terms of success rate.
Novice PCP users had a reasonable success rate ($90\%$) with respect to novice PR2 users ($98\%$). While the PCP and PR2 experts did visibly better than the novice PCP users, they had too few samples to prove a significant difference.

\begin{figure}[!htbp]
\centering
\includegraphics[width=3in]{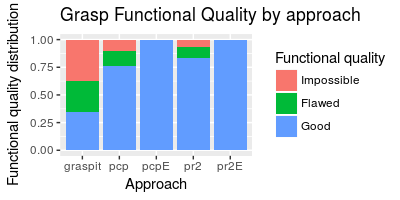}
\caption{Functional grasp quality by approach}
\label{fig:FGC_byApp}
\end{figure}

\paragraph{Grasp Quality Distribution:} Figure \ref{fig:FGC_byApp} shows the distribution of observed events with respect to functional quality for each of the grasping approaches.  

\paragraph{Functional Grasp Success Rate:} 

\begin{figure}[!htbp]
\centering
\includegraphics[width=3in]{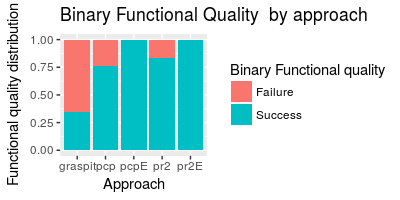}
\caption{Functional Grasp Success rate by approach}
\label{fig:FGsuc_byApp}
\end{figure}

Figure \ref{fig:FGsuc_byApp} shows the distribution of successes vs failures for each of the evaluated approaches with respect to functional quality. In these, \emph{Good} is marked as a \textit{Success}, and as a \textit{Failure} otherwise.
As can be seen, the majority of actions in the methods that involve humans result in a grasp that holds the object in a manner that is appropriate for the objective of the task. GraspIt, focusing only on pure grasp quality has no way to take the high level objective into consideration, and therefore obtains a large quantity of inappropriate grasps.
 
We found a significant difference in functional success rates ($\chi^2_4 = 88.2$, $p<0.0001$).
Table \ref{tab:FGsuc_byApp} shows the result of applying the post-hoc pairwise Chi-squared tests with Bonferroni correction.

\begin{table} [!htbp]
\centering
\begin{tabular}{|c|c|c|c|c|}
\hline         & GRASPIT   & PCP   & $PCP_E$ & PR2  \\
\hline PCP     & <0.0001  & -     & -       & -    \\     
\hline $PCP_E$ & <0.001  & 1.00  & -       & -    \\     
\hline PR2     & <0.0001  & 0.875 & 1       & -    \\ 
\hline $PR2_E$ & <0.0001  & 0.147 & -\textsuperscript{1}    & 0.590\\
\hline 
\end{tabular} 
  \caption{Grasp Success rate per approach} \label{tab:FGsuc_byApp}
\end{table}

\footnotetext{For some comparisons there was not enough data}

What can be seen from the table \ref{tab:FGsuc_byApp}, and Figures \ref{fig:FGC_byApp}, and \ref{fig:FGsuc_byApp}, is that all approaches that involve humans perform better than GraspIt. PCP and PR2 novice users perform similarly with reasonably high success rates ($76.5\%$ and $83.6\%$ respectively). While the PCP and PR2 experts did visibly better than the novice PCP users, they had too few samples to prove a significant difference. No statistical differences were seen between novices and experts for each approach.

\begin{figure}
\centering
\subfloat[][]{\includegraphics[width=3in]{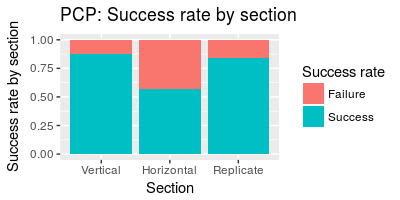}\label{fig:gq_bySec1}}
\\
\subfloat[][]{\includegraphics[width=3in]{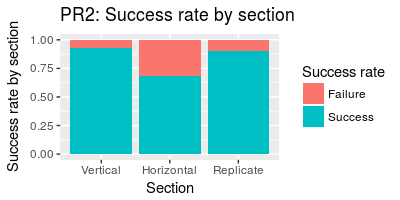}\label{fig:gq_bySec2}}
\caption{ success rates by section in the PCP \protect\subref{fig:gq_bySec1} and in the PR2
\protect\subref{fig:gq_bySec2} experiments}
\label{fig:graspQualbySection}
\end{figure}

It is important to note that while PCP presents a simplified version of the scene, the distribution of functional quality is very similar to that of the live manipulation scenario. Figure \ref{fig:graspQualbySection} shows the distribution of successes and failures with respect to the tasks. 

From our observations and user feedback, we noted that a large percentage of users that initially figured out a grasp for the vertical box scenario attempted a similar grasp for the horizontal box scenario without realizing that the same grasp might not be functionally suitable. This problem appeared in both the PCP and PR2 scenarios, where the results from each section were all statistically equivalent.

This, together with the performance of the expert users indicates that PCP was as good as a live grasping in terms of pure and functional grasping success rates.

\subsubsection{Quantitative Grasp Analysis}

In this section, we analyze the GWS volume and epsilon-distance metrics for each grasp, as well as the time taken to accomplish each task. To do this, we extracted the relative grasp position with respect to each object and exported it to GraspIt. This could be done in the same way in the PCP and PR2 experiments because the pointcloud and object placement was with respect to the exact same coordinate frame: The head-mounted Kinect sensor.

\paragraph{GWS Volume and epsilon-distance: }

\begin{figure}
\centering
\subfloat[][]{\includegraphics[width=1.5in]{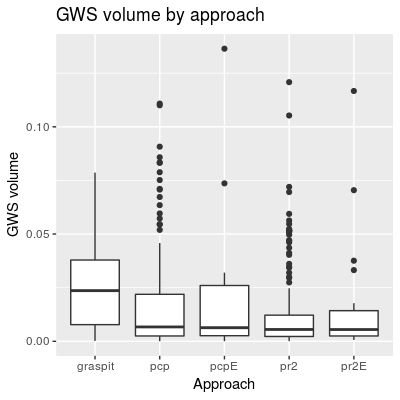}\label{fig:gmet_vol}}
\subfloat[][]{\includegraphics[width=1.5in]{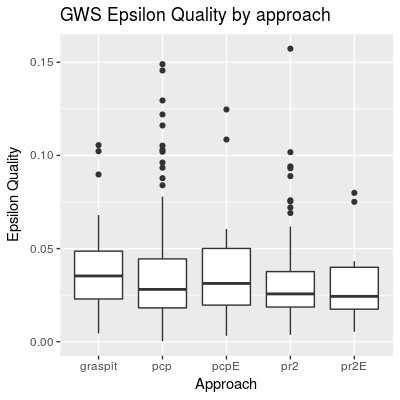}\label{fig:gmet_eps}}
\caption{ GWS volume \protect\subref{fig:gmet_vol} and $\epsilon$-distance
\protect\subref{fig:gmet_eps} by approach }
\label{fig:graspMetrics}
\end{figure}

In Figure \ref{fig:graspMetrics} shows the GWS volume and $\epsilon$-distance metrics for each approach. A statistical difference was found among groups in both the GWS volume ($\chi^2_4 = 40.8$,
$p<0.0001$ )  
and epsilon-distance 
($\chi^2_2 = 10$, $p<0.05$) 
Tables  \ref{tab:gwsvol} and \ref{tab:gwseps} show the results of applying the pairwise Wilcox test to these groups.

\begin{table} [!htbp]
\centering
\begin{tabular}{|c|c|c|c|c|}
\hline         & GRASPIT   & PCP   & $PCP_E$ & PR2  \\
\hline PCP     & <0.0001  & -      & -       & -    \\     
\hline $PCP_E$ & 0.6315   & 1.0000 & -       & -    \\     
\hline PR2     & <0.0001  & 0.8839 & 1       & -    \\ 
\hline $PR2_E$ & 0.0066   & 1.0000 & 1       & 1    \\ 
\hline 
\end{tabular} 
  \caption{GWS Volume by approach} \label{tab:gwsvol}
\end{table}

\begin{table} [!htbp]
\centering
\begin{tabular}{|c|c|c|c|c|}
\hline         & GRASPIT   & PCP   & $PCP_E$ & PR2  \\
\hline PCP     & 0.369  & -      & -       & -    \\     
\hline $PCP_E$ & 1.000  & 1.0000 & -       & -    \\     
\hline PR2     & <0.05  & 1.0000 & 1       & -    \\ 
\hline $PR2_E$ & 0.563  & 1.0000 & 1       & 1    \\ 
\hline 
\end{tabular} 
  \caption{GWS $\epsilon$-distance by approach} \label{tab:gwseps}
\end{table}

Notably, the PCP Expert performed equivalently to the GraspIt Eigengrasp approach for both metrics, and despite being surpassed by GraspIt, novice PCP users achieved equivalent results to all other approaches involving humans.

For PCP, the total number of grasp attempts was $183$, and the number of force-closure grasps was $173$, with a force-closure grasp ``success'' of $94.5\%$. Note that this is not the same as the pure grasp success rate mentioned above ($90\%$) since there are some real grasps that GraspIt evaluates as non-force-closure and some configurations that GraspIt evaluates as valid that in reality slip out of the gripper. 

For the PR2, there were $214$ grasp attempts with $181$ having force-closure. This $84.5\%$ is also different from the observed $98\%$ pure grasp success rate. 

One thing to note is that the magnitudes of both grasp metrics depend on the type of grasp and the robot hand employed. In our case, we have functional-dependent grasps (to complete predefined tasks) and the PR2 parallel gripper (parallel gripper). A multi-fingered robot hand could potentially obtain much higher grasp qualities.

\paragraph{Time:}
We compared grasp timing between the approaches that involved humans. The grasp hypothesis generation using the Eigengrasp planner took $8$ seconds to obtain the $10$ grasp candidates for each case. There are grasp optimization techniques \cite{Ciocarlie2007} that cause it to take up to $\approx 170$ seconds, but we did not employ them.

\begin{figure}
\centering
\subfloat[][]{\includegraphics[width=1.5in]{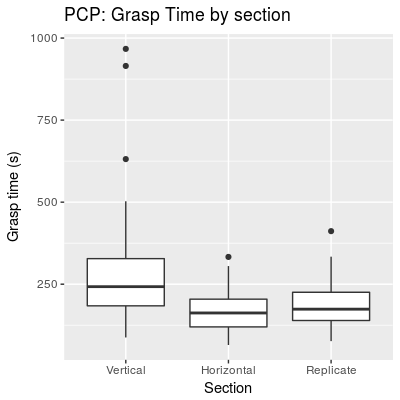}\label{fig:gtimesPCP}}
\subfloat[][]{\includegraphics[width=1.5in]{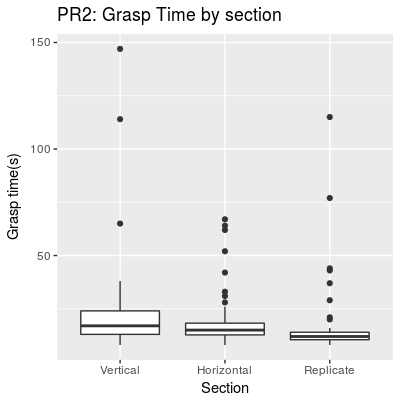}\label{fig:gtimesPR2}}
\caption{ PCP \protect\subref{fig:gtimesPCP} and PR2
\protect\subref{fig:gtimesPR2} grasp times by section}
\label{fig:graspTimes}
\end{figure}

Figure \ref{fig:graspTimes} shows the times to complete tasks within each challenge section for each approach. As can be seen in \ref{fig:gtimesPCP}, novice PCP users achieve substantial speedup from the initial section, with $280$ seconds, to the last, stabilizing at about an average of $184$ seconds, or about $3$ minutes. For the PR2, a slightly smaller speedup was noticed: for the first section, it took them an average of $22$ seconds, and speeding up to $16$ seconds per task for the last section. This points to a rapid increase in abilities under both interaction modes.

For PCP, we found that there was significant difference between sections  ($\chi^2_2 = 34.3$, $p<0.0001$).
Table \ref{tab:graspTimes_pcp} shows the result of applying the pairwise Wilcox test with Bonferroni correction.

\begin{table} [!htbp]
\centering
\begin{tabular}{|c|c|c|}
\hline             & Vertical   & Horizontal    \\
\hline Horizontal  & <0.0001      & -      \\     
\hline Replicate   & <0.0001     & 0.32 \\  
\hline 
\end{tabular} 
  \caption{PCP: Grasp time per section} \label{tab:graspTimes_pcp}
\end{table}

For PR2, we also found a significant difference between sections ($\chi^2_2 = 28.3$, $p<0.0001$).
Table \ref{tab:graspTimes_pr2} shows the result of applying the pairwise Wilcox test with Bonferroni correction.

\begin{table} [!htbp]
\centering
\begin{tabular}{|c|c|c|}
\hline             & Vertical   & Horizontal    \\
\hline Horizontal  & 0.18498      & -      \\     
\hline Replicate   & <0.0001     & <0.001 \\  
\hline 
\end{tabular} 
  \caption{PR2: Grasp time per section} \label{tab:graspTimes_pr2}
\end{table}

Previous systems that employ assisted grasping have achieved relatively good speeds with respect to automatic grasping (if the time to match is around $8$ seconds).  In \cite{weisz2013user}, users achieved an averages of between $86$ and $104$ seconds;  in  \cite{hertkorn2016shared} two strategies were employed, obtaining averages of $28$ and $21$ seconds respectively. We believe that with similar assistance, equivalent speedups can be attained in PCP.

\paragraph{Measured interactions: } In addition to grasp metrics, we analyzed possible interactions between these and the user's spatial reasoning score (SR). 
No correlation was found between SR and the GWS volume and epsilon-distance metrics. However, we found medium effects between SR and grasping duration.

\begin{figure}[!htbp]
\centering
\includegraphics[width=3in]{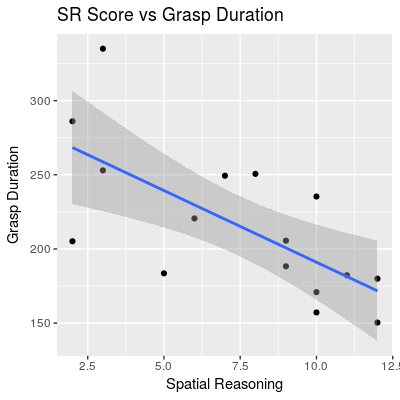}
\caption{PCP only: variation in duration by SR}
\label{fig:SR_vs_graspDur}
\end{figure}

\subparagraph*{Modeling Duration vs SR:}
Figure \ref{fig:SR_vs_graspDur} shows the results of the linear regression where we found a medium effect size (SQ explains $\approx 46\%$ of variation in grasping modeling time), $R^2=  0.4641$,	$F_1,14 = 12.12$,  $p= 0.003664$.

\subsection{Pick-and-Place Handling Results}

The approaches considered were PCP and manual placement using the PR2 robot. Novice users were compared to the expert user(s). For PCP, the total number of participants to complete the minimum number of handling tasks ($3$) was $N_{PCP}^U = 16$. Since we used the same group of people to do the grasp and handling tasks, we only analyze those paths that achieved a successful grasp on the object. A total of $N_{PCP}^H = 166$ novice user handling paths were recorded and compared. For the PR2, we had  $N_{PR2}^U = 18$ participants, who generated $N_{PR2}^G = 214$ handling paths (not all users finished all tasks). 

\subsubsection{Qualitative Handling Analysis}

The PCP handling suggestions were evaluated in the PR2 by using MoveIt to place the end-effector in the indicated pre-grasp waypoint, closing the gripper, and then proceeding with the rest of the handling waypoints. As mentioned above, to evaluate the overall handling path quality (independent of grasping), we looked at three different handling characteristics: collisions, robot configuration problems, and object handling flaws. 

\begin{figure}[!htbp]
\centering
\includegraphics[width=3in]{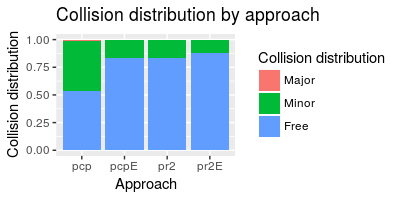}
\caption{Collisions by approach}
\label{fig:HColls_byApp}
\end{figure}

\paragraph{Collisions:} Figure \ref{fig:HColls_byApp} shows the distribution of observed collision events during a handling attempt for each approach.  

Minor collisions include bumping the object with the gripper before grabbing it, and brushing the table or box with the gripper or held object. Major collisions included hitting the object (preventing its grasping), or hitting the table or box with gripper or bottle. These last impacts happen when waypoints cause the gripper or held object to attempt to go \emph{through} the table or box.

To run a Chi-Squared test with success frequencies, we considered major collisions as failures and successes otherwise. No significant differences were found between approaches.

\begin{figure}[!htbp]
\centering
\includegraphics[width=3in]{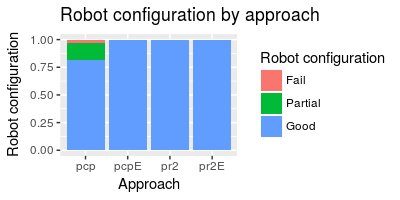}
\caption{Robot Configuration Errors by approach}
\label{fig:HRConfig_byApp}
\end{figure}

\paragraph{Robot Configuration:} Figure \ref{fig:HRConfig_byApp} shows the distribution of observed robot configuration errors during a handling attempt for each approach.  

Partial robot configuration errors cause a single non-crucial waypoint to be skipped because the path would cause a self-collision, or would take some part of the arm out of reach. Fails cause several waypoints or a few crucial ones to be skipped for the same reasons.

To run a Chi-Squared test with success frequencies, we considered partial and good configurations as successes, and failures otherwise.  No significant differences were found between approaches.

\begin{figure}[!htbp]
\centering
\includegraphics[width=3in]{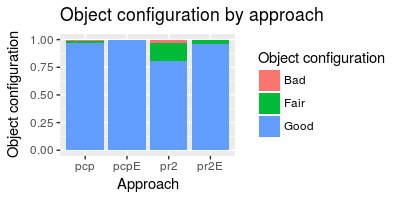}
\caption{Object Configuration Errors by approach}
\label{fig:HOConfig_byApp}
\end{figure}

\paragraph{Object Configuration:} Figure \ref{fig:HOConfig_byApp} shows the distribution of observed object configuration errors during a handling attempt for each approach.  

Fair object configuration errors mean that the object was slightly mishandled. Bad object configurations happen when objects are grossly mishandled in at least one point.

To run a Chi-Squared test with success frequencies, we considered fair and good configurations as successes, and failures otherwise.  No significant differences were found between approaches.

\begin{figure}[!htbp]
\centering
\includegraphics[width=3in]{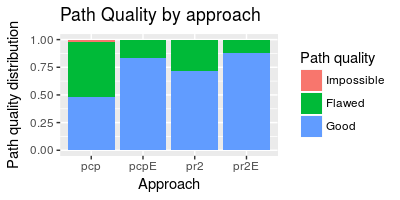}
\caption{Handling quality by approach}
\label{fig:HC_byApp}
\end{figure}

\paragraph{Overall Handling Quality:} Figure \ref{fig:HC_byApp} shows the distribution of observed events during a handling attempt for each approach.  

Flawed overall quality meant that any part of the handling task had a flaw that would not prevent the task from being completed. Attempts counted as impossible would not allow the completion of the task, even with a perfect grasp choice.

To run a Chi-Squared test with success frequencies, we considered \emph{Flawed} and \emph{Good} attempts as successes, and failures otherwise.  No significant differences were found between approaches.

\subsubsection{Quantitative Grasp Analysis}

In this section, we analyze the time and path-distance metrics for each handling, as well as the time taken to accomplish each task. To do this, we extracted the relative waypoint positions with respect to the exact same coordinate frame: The head-mounted Kinect sensor.

\begin{figure}[!htbp]
\centering
\includegraphics[width=3in]{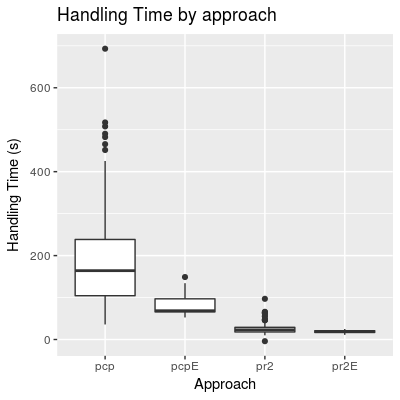}
\caption{Handling time by approach}
\label{fig:Htime_byApp}
\end{figure}

\paragraph{Handling Time:}
The PR2 users were, as expected, faster than those using PCP. We mainly wanted to see how novice and expert users compared under the two approaches, and how PCP users did in comparison to other interface options. Figure \ref{fig:Htime_byApp} shows the distribution of observed handling times for each approach.  Novice PCP users averaged manipulations in $190$ seconds, or about $3$ minutes, while the expert took an average of $84$ seconds, or about $1{\frac{1}{2}}$ minutes. For the PR2, novice users averaged $25$ seconds, with $19$ seconds for the experts.  

\begin{figure}
\centering
\subfloat[][]{\includegraphics[width=3in]{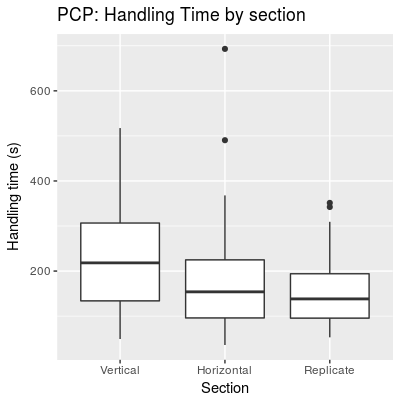}\label{fig:htimesPCP}}
\\
\subfloat[][]{\includegraphics[width=3in]{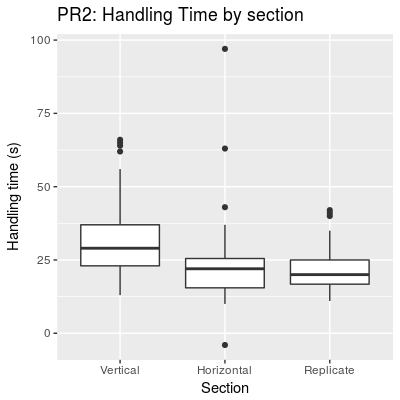}\label{fig:htimesPR2}}
\caption{PCP \ref{fig:htimesPCP} and PR2
   \ref{fig:htimesPR2} handling times by section }
\label{fig:handlingTimes}
\end{figure}

A more detailed view of the handling times can be seen in Figure \ref{fig:handlingTimes}. 
As can be seen in \ref{fig:htimesPCP}, novice PCP users achieve substantial speedup from the initial section, with $227$ seconds, to the last, stabilizing at about an average of $152$ seconds, or about $2{\frac{1}{2}}$ minutes. For the PR2, a similar speedup was noticed: for the first section, it took them an average of $31$ seconds, and speeding up to $21$ seconds per task for the last section. This points to a rapid increase in abilities under both interaction modes.

For PCP, we found that there was significant difference between sections ($\chi^2_2 = 11.5$, $p<0.01$). 
Table \ref{tab:handlingTimes_pcp} shows the result of applying the pairwise Wilcox test with Bonferroni correction.

\begin{table} [!htbp]
\centering
\begin{tabular}{|c|c|c|}
\hline             & Vertical   & Horizontal    \\
\hline Horizontal  & 0.0870      & -      \\     
\hline Replicate   & <0.05     & 1.0000 \\  
\hline 
\end{tabular} 
  \caption{PCP: Handling time per section} \label{tab:handlingTimes_pcp}
\end{table}

For PR2, we also found a significant difference between sections ($\chi^2_2 = 36$, $p<0.0001$). 
Table \ref{tab:handlingTimes_pr2} shows the result of applying the pairwise Wilcox test with Bonferroni correction.

\begin{table} [!htbp]
\centering
\begin{tabular}{|c|c|c|}
\hline             & Vertical   & Horizontal    \\
\hline Horizontal  & <0.001      & -      \\     
\hline Replicate   & <0.001     & 1 \\  
\hline 
\end{tabular} 
  \caption{PR2: Handling time per section} \label{tab:handlingTimes_pr2}
\end{table}

These results indicate that users quickly gain proficiency while maintaining or improving the precision (see path results above). 

\paragraph{Handling Path Distance:}
the absolute magnitude of a handling path distances might not mean anything by itself, since tasks were different and  - for the non-replication tasks - styles varied. We did, however compare path distances between approaches to see if the interface altered this characteristic. It did not.

\begin{figure}[!htbp]
\centering
\includegraphics[width=3in]{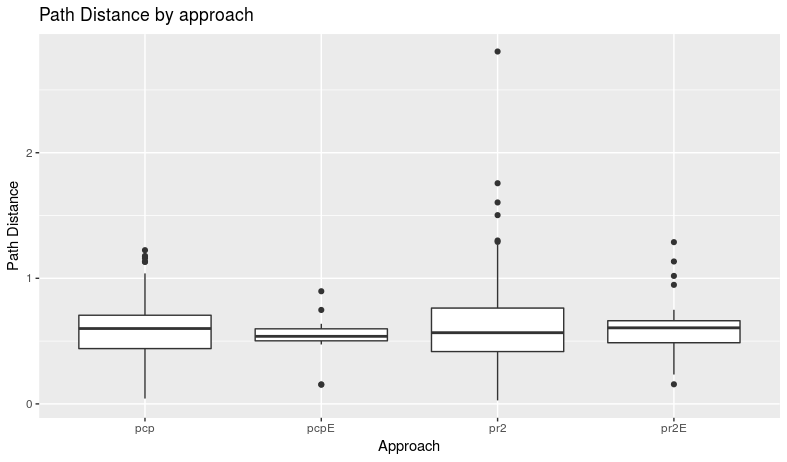}
\caption{Handling distancee by approach}
\label{fig:Hdistance_byApp}
\end{figure} 

Figure \ref{fig:Hdistance_byApp} shows the different path distances for each approach. No significant difference between approaches was found.

\begin{figure}[!htbp]
\centering
\includegraphics[width=3in]{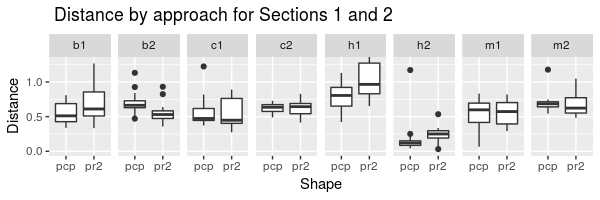}
\caption{Handling Distance for the vertical and horizontal challenges}
\label{fig:Hdistance_vert_hori}
\end{figure}

\begin{figure}[!htbp]
\centering
\includegraphics[width=3in]{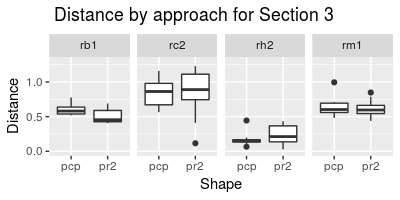}
\caption{Handling Distance for the replication challenge}
\label{fig:Hdistance_rep}
\end{figure}

Figure \ref{fig:Hdistance_vert_hori} shows the distances for each shape per approach for the vertical and horizontal box challenges. Figure \ref{fig:Hdistance_rep} shows the respective distances for the replication challenges. As can be seen, both approaches have very similar results.

\paragraph{Difference between paths (Ribbon Area):}

For the replication challenges, we evaluated the precision with which subjects copied the indicated paths. The ribbon area is the approximate area of the accumulated linear error between the ideal path (the one we prerecorded and showed to subjects for replicating) and the subject-provided one.

\begin{figure}[!htbp]
\centering
\includegraphics[width=3in]{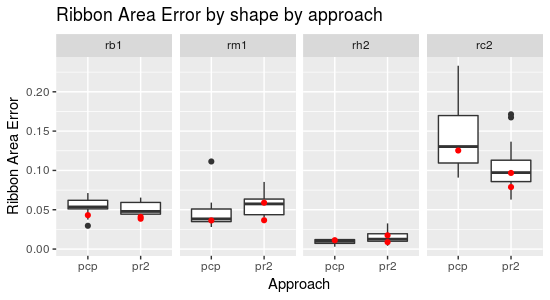}
\caption{Ribbon Area by approach for the replication challenge}
\label{fig:Hribbon_rep}
\end{figure}

Figure \ref{fig:Hribbon_rep} shows the ribbon area error for each shape per approach for the replication challenge. The red dots show the respective errors obtained by the experts.  As can be seen, both approaches have very similar results. For the cases of the bottle and mug, no significant difference was observed. For the handle, when comparing novice PCP users ($M_{PCP}=0.0096 m^2$, $SD_{PCP}=0.0033 m^2$) and novice PR2 users ($M_{PR2}=0.0146 m^2$, $SD_{PR2}=0.0077 m^2$), novice PCP users proved slightly better ($p < 0.05$).  
For the cuboid,  when comparing novice PCP users ($M_{PCP}=0.142 m^24$, $SD_{PCP}=0.0477 m^2$) and novice PR2 users ($M_{PR2}=0.1023 m^2$, $SD_{PR2}=0.030 m^25$), 
novice PR2 users proved slightly better ($p < 0.05$).

\paragraph{Pick-and-Place as a whole (Grasping and Handling):}

\begin{figure}[!htbp]
\centering
\includegraphics[width=3in]{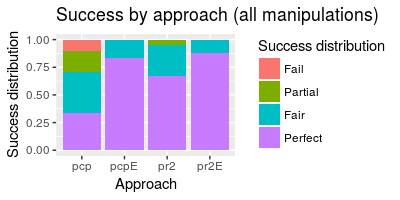}
\caption{Full manipulation quality by approach}
\label{fig:manip_success_byApp}
\end{figure}

When considering the full manipulation task: grasping and handling, the accumulated errors and flaws defined the success rates. Figure \ref{fig:manip_success_byApp} shows the distribution of full manipulation events per approach.

Attempts counted as \emph{Fail} would not allow the completion of the task because of an impossible configuration.
\emph{Partial} overall quality meant that any part of the handling task had a flaw that would prevent the task from being completed, or had several minor flaws in each stage. The attempts marked as \emph{Fair} had a minor flaw that did not affect the overall success of the task. \emph{Perfect} attempts had no issues at all.

To run a Chi-Squared test with success frequencies, we considered \emph{Fair} and \emph{Perfect} attempts as successes, and failures otherwise. We found significant differences in the overall success for each approach ($\chi^2_2 = 51.8$, $p<0.0001$). 
Table \ref{tab:Full_suc_byApp} shows the result of applying the post-hoc pairwise Chi-squared tests with Bonferroni correction. 

\begin{table} [!htbp]
\centering
\begin{tabular}{|c|c|c|c|}
\hline          & PCP          & $PCP_E$ & PR2     \\     
\hline $PCP_E$  & 3.005576e-01 & -     & -    \\     
\hline PR2      & 1.277976e-09 & 1     & -    \\ 
\hline $PR2_E$  & 2.198395e-02        & NaN   & 1.000\\ 
\hline 
\end{tabular} 
  \caption{Grasp Success rate per approach} \label{tab:Full_suc_byApp}
\end{table}

As can be seen in Figure \ref{fig:manip_success_byApp}, in PCP, the number of successful completions was $72.5\%$, a figure similar to the number of grasping attempts that involved no major collisions reported in the work by Leeper et al \cite{leeper2012strategies}. The large percentage of non-perfect executions in PCP is mostly due to the chosen method of value assignment, where we decided to evaluate two or more simple flaws in a manipulation as either \emph{Partial} or \emph{Impossible}. In summary, a third of users were able to indicate a flawless manipulation; $37\%$ had a single minor flaw; $19\%$ had an accumulation of minor flaws that we classified as failures; $10\%$ had outright impossible annotations like grasp misses or slips, or major collisions. In contrast, the PCP and PR2 experts had a $100\%$ success rate (albeit with some minor flaws), and novice PR2 users had a $94.7\%$ overall success rate, out of which $28\%$ had minor flaws.

In the work by Leeper et al \cite{leeper2012strategies}, large decreases in errors were obtained when introducing simple grasping hints. This may be accomplished by simply showing previous successful grasps, as annotations linked to the shape scaffold.
We believe that most or all of the \emph{Partial}, as well as some of the \emph{Fail} marks due to grasping may be eradicated by using this approach. For Handling, one prevalent type of error involve grasping an object in a way that, when completing the handling would put the actual PR2's wrist or forearm sections in contact with some obstacle or object in the work area. A simple modification to aide this would be to extend PCP with a slightly more complete gripper widget.

\section{Discussion \label{sec:discussion}}

\subsection{Overview}

There is a need for extracting accurate representations of a robot's environment that contain noisy data or that have missing information. Given the rise in popularity of point cloud generating depth sensors, methods for extracting accurate object models from point clouds are needed.
While automatic methods for recognizing or reconstructing objects have been developed, there is still a clear need for human input in the recovery of shapes and for assistance in specifying robotic actions.

Annotation of the segmented point cloud data has been shown to help refine object model reconstruction. Additionally, annotation can help guide robot actions or extend the information that an object model can provide. A compact structure that can hold a wide range of annotations is therefore, very useful for robotics.   

The sweep-based scaffold structure is a simple and compact model that can represent a wide range of objects: from synthetic objects composed of primitive shapes, to organic ones with cavities and internal structures. The generalized-cylinder concept is an intuitively simple one to grasp and use as a proxy for more complicated shapes. PCP takes advantage of this fact and is able to provide an interface in which scaffolds can be quickly and accurately created to represent shapes.

Novice users showed that they could quickly learn to use an interface based on the scaffold representation, and that the resulting models had superior shape quality than that of the classic reconstruction method as implemented through KinFu. Additional training showed that an expert user is capable of obtaining shape accuracies that compete with CAD, but that require considerably less time to generate. 

For the reconstruction phase, KinFu-generated point clouds were used, which required some time to create, and that encapsulated imprecisions in their generation that are propagated into the shape modeling with PCP (like the observed effect of point cloud objects looking smaller than their real counterparts). Object modeling does not require these fused point-clouds, but can be done from a single frame that contains enough information for a person to base its model on (which in some cases can be a small fragment of the complete shape). This is a particular benefit of methods that include humans.  

While not currently optimized for that purpose, a further refinement of the modeling tools found in PCP could provide an interface that can be used to crowd-source complex or extensive digitalization efforts. Notably, the merging of user-generated shapes into prototypical ones showed a smaller average shape quality error than that of the average user. This indicated a potential for generating models that incorporate a sort of wisdom-of-the-crowds in the sense that there is a sort of voting scenario where agreement on shape features increase their reliability, and artifacts can be quickly eliminated. Further research is needed to investigate the potential of this approach.  

In terms of grasping and simple manipulation of rigid objects, the object-relative annotations proved as precise and effective as that of live guided manipulation, and proved greatly superior to automatic grasping when the task context was considered. 
Even though the gripper used in the study (PR2 gripper) is rather simple, it was sufficient to compete a wide range of tasks. The representation used in the interface was only composed of the contact surfaces of the gripper, which shows how simple cues are sufficient for a human to achieve complex tasks. On the other hand, the minimalistic gripper representation did not provide the users an idea of collision areas, which resulted in some users inadvertently indicating actions that would cause the gripper to brush or even hit obstructions. A slightly less ``minimalistic'' gripper representation, with at least a simple graphical aspect to the occupancy of the part, could provide sufficient feedback to the user to increase the grasping and handling quality.  While the only gripper used was for the PR2, extending PCL with additional grippers could be done modularly by providing a small file that contains the kinematic chain describing the gripper, and the visualization details for each one. In that sense, it could be a good idea to adopt a standard gripper description format, like the one used by GraspIt!.
While the focus of this study is not fine-level grasping, the grasp annotation to the scaffold structure could be coupled with grasp-adjustment methods \cite{hsiao2010contact} to achieve even higher quality grasps. In addition, the grasping and handling annotations (represented by object-relative pose waypoints) could also be used to plan higher-level tasks. Even though this was not the focus of the thesis, the saved object models and their saved modes of use could be used as action primitives in the construction of more complex behaviors.

\subsection{Limitations}

Some shapes are still beyond the scope of this approach. This has to do with the time required to accurately represent complex topologies or high feature repetition. Some modeling languages compensate these problems by incorporating scripted part generation routines, that can be specified programmatically, or by using analytical expressions. Currently, no such features exist in PCP, but given the simplicity of the scaffold proxy, it would not take much to extend it with such capabilities.

Some grasping issues arose from the specific gripper we used for the experiments. 
Grasping the lid of the travel mug was difficult to indicate with no live-control. This is due to the relative size of the gripper to the lid. While grasps were successfully placed by the expert, this item was discarded from the user study because of this factor. As mentioned above, grasp-adjustment methods can be used to deal with this low-level problem.

\subsection{Summary}

The scaffold structure is a simple and expressive representation that was shown to help in the modeling of a wide range of shapes, as well as in the indication of three typical robotics tasks: grasping, pick-and-place, and articulated-object manipulation. 

The DARPA Robotics Challenge emphasized the benefits of object models in indicating robotic models. It also showed the usefulness of having a way of quickly creating said models, and add them to the manipulation capabilities of the system. The teams that incorporated these models into their shared-control strategies were among the most successful ones.

On a high-level perspective, it is easy to see the benefits of the object-relative annotation approach to shared-control when considering that often the person that makes or knows the object (the \emph{object-maker}) has a better idea of how it should be manipulated than the operator of the robot (\emph{the robot-mover}). 
A simple tool, and a simple primitive should be available for \emph{object-makers} to indicate these preferred modes of use.

\section{Conclusions and Future Work \label{sec:conclusions}}

Shared control with Human-in-the-Loop is very useful, and in some cases, required, to get a robot to complete certain challenges. In a wide range of scenarios, continuous teleoperation has been superseded by a variety of shared-control mechanisms that explore different methods of splitting the workload that a human and a robot must complete.  

While many strategies for autonomously planning parts of the interaction exist, object-relative annotations definitively have shown to provide excellent results. For maximum utility, these annotations should be saved in a simple, extensible and reusable container.

The shape-based annotation container that we present, called Point Cloud Scaffold was shown to provide enough expressive power to represent a wide variety of shapes that could be encountered by a service robot in an indoor environment.
In addition, the annotations that can be attached to this structure benefit from the scaffold design itself: The hierarchical nature of the object (whole $\rightarrow$ part $\rightarrow$ scaffold $\rightarrow$ slice $\rightarrow$ handle), as well as the parameters that define the structures (slices have planes with normals, and scaffolds have sweep-axes). 

The object modeling experiment shows that novice users can quickly learn to use these shape proxies to represent a wide range of objects with high accuracy, and that, given more experience, these objects can compete with CAD in terms of precision, while requiring a fraction of the time to generate.

The Pick-and-Place experiment showed that object-relative grasping and handling annotations can be very effective and easy to indicate, and that once created, their reuse is a simple matter of adjusting the waypoints. The grasp and pick-and-place annotations were tested on the PR2 and a high success rate was achieved, even for novice users.

\subsection{Future Work}

In the following sections, the possible next steps of this work are detailed.

\subsubsection{Prototype Scaffolds and Mean Shapes}

We would like to further investigate the potential of the mean shapes obtained from merging multiple individual user submissions. The steps would include the following:

\begin{enumerate}
\item Design a method for merging the PCSs themselves - using geometric properties - into a \emph{prototype scaffold}.  These annotation styles could be classified through the characteristics of the scaffolds themselves, like the number of parts or the sweep-axis direction. 
If more than one set of annotation styles is detected, obtain a \emph{prototype scaffold} for each.
\item Evaluate the \emph{prototype scaffold} in terms of shape reconstruction using the same metrics as in the user study presented in this thesis.
\end{enumerate}

An advantage of obtaining a ``running average'' of user scaffolds (in the shape of the prototype scaffold) would be that an immediate shape hint would be available for novice users to base the shape on. If multiple styles exist, then multiple starting-points could be provided. This would have an immediate effect on reconstruction speed. A possible application could involve a continually refined system for crowdsourcing object reconstruction where modeling suggestions, as starting points, become available as more people participate.

\subsubsection{Prototype Grasps}

A consequence of merging scaffolds for reconstruction could be that the same might be done for grasping. A series of steps for doing so with the PCS structure would involve the following steps:

\begin{enumerate}
\item Design a method to transfer grasp annotations from an individual PCS to the \emph{prototype scaffold}, while retaining or improving the grasp quality.
\item Design a method to cluster grasp annotations into different groups - using geometric and functional relationships - and merge each cluster into a \emph{prototype grasp}. 
\item Evaluate the \emph{prototype grasp} in terms of object grasping using the same metrics as in the user study.
\end{enumerate}

This would have an immediate impact on annotation speed as well. An added factor is the functional dimension. Prototype grasps could represent clusters that not only differentiate preference, but objective; e.g. choosing to grab a mug through the handle for pouring, or from the top, for stacking.

\subsubsection{Scaffold Database and Annotation Transference}

One benefit of the scaffold structures are that they can be easily classified through their structural properties into a database of object proxies that link to example underlying clouds and output meshes. In addition to the direct geometry, topological and functional properties can be used to further link the scaffolds within the database. Further annotation levels (like articulation or feature extraction) may generate additional parameters by which to link objects into classes and uses.

A related feature would be the transference of annotations between instances of the same class. An example would be to transfer the joint constraints to open a jar by turning the lid from one scaffold to another just by identifying that both fall under the same category: jars with screw lids, or in geometric/functional terms: "containers with a screw joint to a link covering its concavity".

A possible application of such a database may be the quick search and insertion of a scaffold into a scene point cloud. This feature, in conjunction to transference of annotations, could allow for a rapid geometric and functional annotation of multiple objects in an environment where a robot must operate. This could be done in a crowd-sourced manner for further speed gains.

\subsubsection{Annotations For Articulated or Deformable Objects}

An immediate extension of this work is to add articulation annotations that constrain the motion between the parts of a complex (multi-link) object. Any actuation by a robotic hand can be therefore constrained to the desired planes and exes of articulation.
One possible level of annotations could relate to the reaction to force for each of the parts of the scaffold. The geometric properties of the scaffold re such that the occupied volume of the objects can be reshaped using a tetrahedral mesh and spring-based parametric annotations for the edges to denote resistance to force.

\bibliographystyle{ACM-Reference-Format}
%\bibliography{modelingPaper-bibliography} 
\bibliography{modelingPaper} 
\end{document}